\documentclass[useAMS,usenatbib]{mnras}
\usepackage[english]{babel}
\usepackage[fleqn]{amsmath}
\usepackage{amsfonts}
\usepackage{txfonts}
\usepackage{graphicx}
\usepackage{amssymb}
\usepackage[T1]{fontenc}
\usepackage{ae,aecompl}
\usepackage{fancyhdr}
\usepackage{multicol}
\usepackage{layout}
\usepackage{graphicx}
\usepackage{multirow}
\usepackage{color}
\usepackage{times}
\usepackage[outdir=./]{epstopdf}

\newif\ifAMStwofonts
\AMStwofontstrue

\def\karl#1{{ #1}}
\def\newkarl#1{{ #1}}
\newcommand{\dd}{\mathrm{d}}

\title[Investigating scalar-tensor-gravity]{Investigating scalar-tensor-gravity with statistics of the cosmic large-scale structure}
\author[R. Reischke, A. Spurio Mancini, B.M. Sch{\"a}fer and Ph.M. Merkel]{Robert Reischke$^{1}$\thanks{e-mail:
reischke@stud.uni-heidelberg.de}, Alessio Spurio Mancini$^2$, Bj\"orn Malte Sch\"afer$^{1}$ and Philipp M. Merkel$^3$\\
$^1$ Zentrum f{\"u}r Astronomie der Universit{\"a}t Heidelberg, Astronomisches Rechen-Institut, Philosophenweg 12, 69120 Heidelberg, Germany\\
$^2$ Institut f\"ur Theoretische Physik, Universit\"at Heidelberg, Philosophenweg 16, 69120 Heidelberg, Germany\\
$^3$ Zentrum f{\"u}r Astronomie der Universit{\"a}t Heidelberg, Institut f{\"u}r Theoretische Astrophysik, Philosophenweg 12, 69120 Heidelberg, Germany}

\begin{document}

\pagerange{\pageref{firstpage}--\pageref{lastpage}} \pubyear{}
\maketitle

\label{firstpage}

\begin{abstract}
Future observations of the large-scale structure have the potential to investigate cosmological models with a high degree of complexity, including the properties of gravity on large scales, the presence of a complicated dark energy component, and the addition of neutrinos changing structures on small scales. Here we study Horndeski theories of gravity, the most general minimally coupled scalar-tensor theories of second order. While the cosmological background evolution can be described by an effective equation of state, the perturbations are characterised by four free functions of time. We consider a specific parametrisation of these functions tracing the dark energy component. The likelihood of the full parameter set resulting from combining cosmic microwave background primary anisotropies including their gravitational lensing signal, tomographic angular galaxy clustering and weak cosmic shear, together with all possible non-vanishing cross-correlations is evaluated; both with the Fisher-formalism as well as without the assumption of a specific functional form of the posterior through Monte-Carlo Markov-chains (MCMCs). Our results show that even complex cosmological models can be constrained and could exclude variations of the effective Newtonian gravitational coupling larger than 10\% over the age of the Universe. In particular, we confirm strong correlations between parameter groups. Furthermore, we find that the expected contours from MCMC are significantly larger than those from the Fisher analysis even with the vast amount of signal provided by stage IV experiments, illustrating the importance of a proper treatment of non-Gaussian likelihoods and the high level of precision needed for unlocking the sensitivity on gravitational parameters.\newline
\end{abstract}


\section{Introduction}
Combination of different cosmological probes such as type Ia Supernovae \citep[SNIa, e.g.][]{Perlmutter1997, Riess1998, Perlmutter1999, Riess2004,Riess2007}, the angular power spectrum of the cosmic microwave background (CMB) anisotropies \citep[e.g.][]{Hinshaw2013,Planck2016_XIII} and of galaxy clustering \citep[e.g.][]{Cole2005} led to the conclusion that the Universe is expanding in an accelerated fashion. The concordance cosmological model, $\Lambda$CDM, is based on a description of space time by general relativity (GR) with a non-zero cosmological constant $\Lambda$, filled with cold dark matter (CDM), photons and baryons. So far, the $\Lambda$CDM model has been consistent with almost all observations.

While SNIa probe the cosmological background model only and CMB experiments are sensitive to early time structure formation, upcoming surveys of the large-scale structure will provide exquisite data of the perturbed Universe. Even if the cosmological background seems to be very well described by the $\Lambda$CDM model, perturbations on this background could still be different. More importantly, general relativity has been tested on non-cosmological scales and in the weak field limit only \citep[see][for reviews,]{heavens_model_2007, jain_observational_2008, bertschinger_distinguishing_2008, Berti2015}, with inconclusive answers on tensions of the data with $\Lambda$CDM \citep{giannantonio_new_2010, dossett_constraints_2015}. More recently it also has been found to be consistent with binary black hole mergers \citep{Ligo2016} and neutron star mergers \citep{Abbott2017b}. The latter showed that the expansion speed of tensorial modes is equal to the speed of light. In the standard paradigm one assumes that the physical laws are also valid on cosmological scales. Even for general relativity this corresponds to an extrapolation passing many orders of magnitude. Especially general relativity is applied in the strong curvature regime in cosmology since the scales of interest are often of comparable size as the horizon.

From this discussion it should be clear that general relativity needs to be tested on cosmological scales {\citep{lue_differentiating_2004, laszlo_non-linear_2007, kunz_dark_2007, koyama_cosmological_2016, joyce_dark_2016, white_marked_2016}}. In order to test it, observable consequences of modifications to general relativity need to be established and compared to observational results \citep{koyama_structure_2006}. Those modifications of general relativity lead to very different phenomena \citep[see][for a review]{Clifton2012} and influence the background expansion as well as the growth of structures \citep{zhao_searching_2009, kobayashi_evolution_2010}. Furthermore, these models are highly degenerate with (clustering) dark energy models \citep[e.g.][]{Copeland2005}, which can as well reproduce very different expansion histories and structure growth. This is already evident from the structure of the field equations of GR: modifications in the gravitational sector can be interpreted as a modifications in the matter sector and can thus be treated as some kind of dark energy component \citep{Battye2012, Battye2013}. Consequently, cosmological tests require exquisite data with high statistical significance in order to have strong enough signals to constrain nuisance, cosmological, gravitational and particle physics parameters \citep{Clifton2012, boubekeur_current_2014}.

In the next decade we expect a huge step forward in large scale surveys dubbed stage IV experiments \citep{albrecht_report_2006}. These are in particular the \textsc{Euclid} mission \citep{Laureijs2011} or \textsc{LSST} \citep{LSST2009} for galaxy clustering and weak gravitational lensing, {\karl{CMB stage IV experiments \citep{CMBS4}}}, or surveys of the distribution of neutral hydrogen through its hyperfine transition with the \textsc{Square Kilometer Array} (SKA) \citep{Maartens2015}. Operating at very different redshift and scales, cross-correlation between those experiments will provide significant information about the underlying gravity model. Additionally these missions will also deliver tests of fundamental physics such as the mass scale of the neutrinos \citep{pettorino_neutrino_2010, Ribera2014}. This is of particular importance as the neutrino masses are highly degenerate with modifications of the gravity sector \citep{Baldi2014, Baldi2016, merkel_parameter_2017}. Being very sensitive to the mass of the neutrinos, CMB experiments will yield complementary information to weak lensing surveys or galaxy clustering which are sensitive to the gravitational slip or the modified Poisson equation respectively.

Having a high dimensional parameter space, advanced sampling techniques are required, even for forecasting to yield  conservative errors on the cosmological parameters. The inference process itself is carried out with maximum likelihood methods and the best fit parameter is given by the maximum likelihood estimator (MLE). By virtue of Bayes' theorem the likelihood can be cast into a posterior distribution for the cosmological parameters. The statistical errors on the parameters are given by confidence intervals calculated from the Monte-Carlo Markov-Chain (MCMC).

A typical forecasting tool is the Fisher-matrix which approximates the posterior distribution as a Gaussian. This assumption is only exact for linear model parameters, and in the general, non-linear case one can at least expect to find approximate Gaussian likelihoods as soon as the data constrain the possible parameter values to fall into a sufficiently small range of values inside which a linearisation of the dependence between the model prediction and the choice of the parameter value is applicable. In this limit the statistical errors predicted by the Fisher-matrix formalism should correspond well to the true variances and the Cram{\'e}r-Rao inequality turns into an equality. The amount of data for cosmological measurements is, however, limited in a natural way, consequently the Cram\'{e}r-Rao bound can never turn into an equality: unlike in other branches such as particle physics, cosmological parameters can only be determined simultaneously, and the hierarchy\footnote{With hierarchy we refer to the different accuracy at which cosmological parameters can be measured, which is limited by the accessible information and the sensitivity of the probes.} of parameters typical for cosmology requires exquisite precision on some fundamental parameters (for instance $\Omega_\mathrm{m}$ and $\sigma_8$) before meaningful constraints on subtle effects such as modifications of gravity or dark energy can be derived. 

In this work we study the constraining power of future experiments on modified gravity including several other cosmological parameters as well as neutrino masses using both a Fisher analysis and Monte-Carlo Markov chains. In particular, we employ affine invariant, parallelised sampling to investigate the shape and orientation of likelihoods in a 17-dimensional parameter space composed of parameters governing the background dynamics, changes in gravity, and astrophysical parameters \citep{foreman-mackey_emcee:_2013,santos_bayesian_2016}.

Ideally, one would like to be as general as possible when investigating gravitational theories, that is covering the largest possible theory space. From an observational point of view there are phenomenological parameterisations such as the $\mu$, $\gamma$ parametrisation \citep{2015arXiv150201590P}. It is, however, not clear to which theories this parametrisation is related. On the other hand there exists a very wide range of gravitational theories and it is necessary to bundle them in different classes. At the background level it is clear that an effective equation of state, $w(a)$, of the dark energy describes the evolution of the metric completely, because density and pressure as a function of time are the only degrees of freedom a relativistic fluid can have under the assumption of the Friedmann symmetries. 

Dark energy is usually described using the first two terms of the Taylor expansion $w(a) = w_0 +(1-a)w_\mathrm{a}$ \citep{Chevallier2001}. On a perturbative level the description is more complicated. A common method to deal with the perturbations is the effective field theory (EFT) approach. Here, one sets up an action for the perturbations containing all operators obeying the symmetries of the background cosmology as well as the properties of the underlying effective dark energy description. The coefficients are then functions of time only and fully describe the perturbations at a linear level.

Here, we particularly focus on Horndeski theories of gravity \citep{Horndeski1974, Nicolis2009, Deffayet2011} which represent the most general minimally coupled scalar-tensor theory of gravity. These theories are free from ghost-like degrees of freedom, since the field equations do not contain derivatives higher than second order. Especially, we will use the \textsc{hiCLASS} code \citep{Zumalacarregui2016} to calculate the perturbations in the EFT framework on a background described by an effective equation of state which was introduced for the Horndeski class in \citet{Bellini2014}. It allows to describe the linear perturbations by four functions of time only. Due to measurements of gravitational wave signals \citep{Ligo2016, Abbott2017a, Abbott2017b} the freedom in those functions can be reduced even further since the tensor speed excess $\alpha_\mathrm{T}$ is basically fixed to its General Relativity value. In this context \citet{Ezquiaga2017,Creminelli2017} as well as \citet{Amendola2017,Sakstein2017} discussed the particular implications on Horndeski theories, showing that the scalar can be coupled only conformally to the curvature.
Furthermore, the kineticity does not influence cosmological observables significantly and it is basically uncorrelated with the remaining free functions. 

This leaves two free degrees of freedom in this class of models, the running of the Planck mass $\alpha_\mathrm{M}$ and the braiding $\alpha_\mathrm{B}$, describing the clustering of the dark energy component on small scales.

We will investigate the constraining power of combined observations of CMB temperature and polarisation anisotropies, CMB lensing, galaxy clustering and cosmic shear on the remaining two degrees of freedom and the neutrino mass scale. 

The structure of the paper is the following: In \autoref{sec:gravity} we summarise the basic properties of Horndeski theories of gravity. Then, in \autoref{sec:data} we introduce the cosmological probes used. In \autoref{sec:statistics} the statistical methods are described. We present the results in \autoref{sec:results} and summarise in \autoref{sec:conclusion}.

\section{Scalar-tensor theories of gravity}\label{sec:gravity}
The most general scalar-tensor theory of gravity \citep{Horndeski1974} has the following Lagrange density:
\begin{equation}\label{eq:Horndeski_Lagrangian}
\mathcal{L} = \sum_{i=2}^5 L_i\left[\phi,g_{\mu\nu}\right] + L_\mathrm{m}\left[g_{\mu\nu},\psi\right],
\end{equation}
such that the action is $S = \int\dd^4x\sqrt{-g}\mathcal{L}$, where $\dd^4x\sqrt{-g}$ is the canonical volume form, $\phi$ the scalar field, $g_{\mu\nu}$ the metric and $\psi$ some collection of matter fields. The individual terms in the Lagrange density are given by
\begin{equation}
\begin{split}
{L}_2 &= G_2 (\phi, X), \\
L_3 &= -G_3(\phi, X) \Box \phi,  \\
{L}_4 &=  G_4 (\phi, X) R + G_{4X}(\phi, X) \left[ (\Box \phi)^2 - \phi_{;\mu\nu} \phi^{;\mu\nu} \right],\\
{L}_5 &= G_5 (\phi, X) G_{\mu\nu} \phi^{;\mu\nu}  \\ 
&- \frac{1}{6}G_{5X} (\phi, X) \left[ (\Box \phi)^3 + 2 \phi_{;\mu}^{\nu} \phi_{;\nu}^{\alpha} \phi_{;\alpha}^{\mu} - 3 \phi_{;\mu\nu} \phi^{;\mu\nu} \Box \phi \right].
\end{split}
\end{equation} 
Here we labelled the kinetic term of the field $X \equiv - \nabla_\nu \phi\nabla^{\nu}\phi/2$. The four functions $G_j$ and $K=G_2$ can in principle be chosen freely and characterise the theory completely. Note that this theory is only minimally coupled to matter through the canonical volume form, extending this particular coupling would go beyond the class of Horndeski gravity. Covariant derivatives are denoted by semicolons.

In \citet{Bellini2014} it was shown that the evolution of linear perturbations in Horndeski theories can be completely characterized by free functions depending on time only:
\begin{equation}
\begin{split}
M_*^2 & =\  2\left(G_4 -2X G_{4X} + XG_{5\phi}-\phi HXG_{5X}\right)\, , \\
HM_*^2\alpha_\mathrm{M} & \equiv \frac{\mathrm{d}M_*^2}{\mathrm{d}t}\, , \\
H^2M^2_*\alpha_\mathrm{K} & \equiv \ 2X\left(K_X + 2XK_{XX} - 2G_{3\phi} - 2XG_{3\phi}\right)\\
& +12\dot\phi XH\left(G_{3X}+XG_{3XX} -3G_{4\phi X} - 2XG_{4\phi XX}\right) \\
& +12X H ^2\left(G_{4X} +8XG_{4XX} + 4X^2G_{4XXX}\right) \\
&-12XH^2 \left(G_{5X}+5XG_{5\phi X} +2X^2G_{5\phi XX}\right) \\
&+14\dot\phi H^3\left(3G_{5X} +7XG_{5XX} +2X^2G_{5XXX} \right)\, , \\
HM_*^2\alpha_\mathrm{B} & \equiv 2\dot\phi\left(XG_{3X} -G_{4\phi} -2XG_{4\phi X} \right)\\
& +8XH\left(G_{4X}+2XG_{4XX} -G_{5\phi}-XG_{5\phi x}\right) \\
& +2\dot\phi XH^2\left(3G_{5X} +2XG_{5XX} \right), \\
M_*^2\alpha_\mathrm{T} & \equiv 2X\left(2G_{4X} -2G_{5\phi} -(\ddot{\phi}-\dot\phi H)G_{5X}\right).
\end{split}
\end{equation}
Here $M_*$ is the Planck mass and $\alpha_\mathrm{M}$ describes its time evolution, it has thus direct implications on the gravitational interaction via the Poisson equation. $\alpha_\mathrm{K}$ describes the kinetic energy and is thus largely unconstrained by observations, since there is no direct influence on any observable. In contrast, the braiding $\alpha_\mathrm{B}$ describes how $\phi$ itself mixes with the scalar perturbations of the metric. Lastly $\alpha_\mathrm{T}$ basically describes the propagation speed of tensorial modes and how it differs from normal null geodesics. Therefore, both $\alpha_\mathrm{M}$ and $\alpha_\mathrm{T}$ affect the propagation of gravitational waves, they can be constrained rather well by non-cosmological experiments \citep[e.g][]{Lombriser2015,Velten2017}. Clearly, these functions restrict the evolution of perturbations in scalar-tensor theories of gravity which can be parametrised by only a few numbers. A common choice for a parametrisation would be
\begin{equation}\label{eq:parametrization_de}
\alpha_i = \hat{\alpha}_i \Omega_\mathrm{DE} + c_i,
\end{equation}
since in such a way the modifications track the accelerated expansion of the Universe. Quite obviously, this approach is idealised; however it gives us a good idea what can be learned from data in these very general models. For a more detailed discussion on these topics we refer to \citet{Linder2016} and \citet{Alonso2017}.
In Newtonian gauge the linear perturbation equations in Fourier space are given by
\begin{equation}\label{eq:Poissoneq}
k^2\Phi_k = -\frac{3H_0^2\Omega_\mathrm{m}}{2a}\delta_k\mu(k,a)
\end{equation}
and
\begin{equation}\label{ea:bardeenratio}
\frac{\Phi_k}{\Psi_k} = \gamma(k,a),
\end{equation}
where the subscript $k$ denotes the corresponding wave-number in Fourier space. 
It should be noted that we do not refer to any particular parametrization here {\karl{with which Eq. (\ref{eq:Poissoneq}) would be valid only in the quasi-static regime. The functions $\mu$ and $\gamma$ describe the direct mapping between the density and the potential fluctuations and the two potential fluctuations respectively. We will keep this notation throughout this paper such that one could adapt everything to particular parametrizations in $\mu$ and $\gamma$ with keeping in mind that the results derived here do not make use of the quasi-static approximation.}}

Many modified gravity models are equipped with a screening mechanism such that at small scales or high density regions general relativity is recovered {\citep{babichev_introduction_2013, Joyce2015}}. This mainly helps the theories to survive local tests of gravity in the solar system. Screening effects become important at small cosmological scales, such that the effect of modified gravity is suppressed with respect to linear theory \citep{Barreira2013,Li2013,Winther2015}. So far non-linear structure formation in modified gravity scenarios is only done for very specific models, but not for such a general class as Horndeski, even though a lot of the theories covered by Horndeski theories have an effective screening mechanism which is, however, only present at the non-linear level. In this work we will use linear predictions and then apply the halo model correction to the linear predicted power spectrum. We justify this choice by a phenomenologically chosen screening scale $k_\mathrm{s}$ imposed on the $\alpha$-functions \citep{Alonso2017}:
\begin{equation}\label{eq:screeningalpha}
\alpha_i \to \alpha_i\exp\left[\left(-\frac{k}{2k_\mathrm{s}}\right)^2\right].
\end{equation}

\section{Cosmological probes}\label{sec:data}
Each data set is given by a collection of spherical harmonic modes $\{ a^X _{\ell m}\}$ where $X$ labels the probe considered. In this work we will focus on CMB primary anisotropies, i.e. temperature fluctuations ($T$) and two polarisation modes ($E,\ B$), the CMB lensing signal deflection field ($D$), tomographic galaxy clustering ($g_i$) and cosmic shear ($\gamma_i$). If the modes follow a Gaussian distribution they can be fully described by their angular power spectra. Since the modes of different data sets are not independent, the cross-spectra will not vanish and carry valuable cosmological information. On the other hand the correlation between the different data sets reduces the independence of the modes resulting into less information. 

In this section we will discuss all relevant spectra together with their respective noise properties.
Throughout this section we will always use the Limber projection \citep{Limber1953} whose validity was discussed in great detail in \citet{Kitching2017}. Throughout this section we will use $\ell' = \ell + \frac{1}{2}$, as a better approximation for projecting spectra.

\subsection{Structure growth}\label{subsec:structuregrowth}
As outlined in \autoref{sec:gravity}, modified gravity mainly influences the growth of structures, i.e. the metric perturbations. For evolving the perturbations we use the Boltzmann code \texttt{hi\char`_class} \citep{Zumalacarregui2016} which is an extension of \textsc{CLASS} \citep{Lesgourgues2011} for scalar-tensor theories of gravity. The code evolves initial conditions set by perturbations forward in time using linear equations only. At small scales and low redshifts the perturbations become non-linear. On these scales the treatment of the Boltzmann codes breaks down.   
For the CMB primary anisotropies this effect is small, since the only non-linear contributions come from the integrated Sachs-Wolfe effect which has most of its power on large angular scales. 

On the contrary cosmic shear and galaxy clustering will acquire the majority of their signal from non-linear scales. It is therefore necessary to model the non-linear scales carefully to exploit the full power of future surveys. A lot of effort has been put into understanding the non-linear evolution theoretically in standard cosmological models in both analytical models \citep[e.g][]{ZelDovich1970, Buchert1992, Bouchet1995, Bernardeau2002, Cooray2002, Hilbert2011, Bartelmann2014, Bartelmann2017}, and through simulations \citep[e.g.][]{Smith2003,  Heitmann2010}. In modified gravity theories non-linear models are more difficult due to the modified Poisson equation and so far no common description of non-linear scales in general tensor-scalar theories \citep[e.g.][]{Hu2007, Zhao2011, Casas2017} exists, and only restricted simulations of non-linear structure formation within general relativity \citep{adamek_general_2013} or certain types of $f(R)$-gravity \citep{achitouv_imprint_2015} are available. In this work we take care of non-linearities by using the Halofit model \citep{Smith2003, Takahashi2012, Mead2015}. It is thus assumed that the effect of modified gravity is negligible on small scales and that general relativity yields an adequate description \citep{Hu2007}. For Horndeski cosmologies this is not true in general, since there is no screening evolved. It is therefore necessary to include the screening by hand as described in Eq. (\ref{eq:screeningalpha}). 

Phenomenologically, the effect of screening can be better understood by looking at the effective Newtonian coupling, the gravitational slip and the linear growth factor:
\begin{equation}
\begin{split}\label{eq:screening}
\mu(k,a)\to & \ \mu_\mathrm{GR} + \mu_\mathrm{MG}(a,k) W(k;k_\mathrm{s}) \\
\gamma(k,a)\to & \ \gamma_\mathrm{GR} + \gamma_\mathrm{MG}(a,k) W(k;k_\mathrm{s}) \\
D_+(k,a)\to & \ D_\mathrm{+GR}(a) + D_\mathrm{+MG}(a,k) W(k;k_\mathrm{s}).
\end{split}
\end{equation}
Here $W(k,k_\mathrm{s}) \propto \exp(-(k/k_\mathrm{s})^2)$ is a low-pass filter to ensure that on scales $k>k_\mathrm{s}$ general relativity is recovered. An usual value for the screening scale would be $k_\mathrm{s} \approx 0.1 h/\mathrm{Mpc}$. 
In the limit of general relativity, $\mu$ and $\gamma$ are unity, while the linear growth factor $D_+(a)$ can be recovered from
\begin{equation}
\frac{\mathrm{d}^2}{\mathrm{d}a^2}D_+(a) + \frac{1}{a}\left(3+\frac{\mathrm{d}\mathrm{ln}H}{\mathrm{d}\mathrm{ln}a}\right)\frac{\mathrm{d}}{\mathrm{d}a}D_+(a) -\frac{3}{2a^2}\Omega_\mathrm{m}(a)D_+(a) =0.
\end{equation}

Furthermore, baryonic effects such as feedback mechanisms become important on small scales, since the dark matter particles are coupled to the baryons by gravity. \citet{Schneider2015} proposed a fitting formula able to describe the effect on the matter power spectrum for a variety of feedback models. The fitting formula depends on several parameters, for instance the average halo mass, $M_\mathrm{c}$, below which most of the gas is ejected, and the maximum scale, $\eta_\mathrm{b}$, up to which the suppression is active.

\subsection{CMB primary anisotropies}\label{sec:CMB}
Maps of the CMB temperature and polarisation contain various kind of primary anisotropies which are mainly related to the potential landscape at the last scattering surface. Foregrounds induce secondary anisotropies, which, however, can be removed very well as they destroy the thermal CMB spectrum. The cleaned maps of CMB data are given in spherical harmonics of $P=T,E,B$ with some instrumental noise $n$, with root mean square $\sigma^2_P$. Instrumental noise CMB experiments can be modelled by a Gaussian beam with width $\theta_\mathrm{beam}$. Additionally one adds white noise to find the noise covariance to be \citep{knox_determination_1995}
\begin{equation}
N^{P}(\ell)\equiv \langle n^{P*}_{\ell m}n^{P'}_{\ell m}\rangle = 
\theta^2_\mathrm{beam}\sigma^2_P\exp\left(\ell(\ell+1)\frac{\theta_\mathrm{beam}^2}{8\mathrm{ln}2}\right)\delta_{PP'}\,.
\end{equation}
diagonal in $P$ as the noise of different maps is uncorrelated, likewise one obtains diagonality in $\ell$ and $m$ for full-sky observations of statistically homogeneous and isotropic fields.

Consequently, the angular power spectrum is given by
\begin{equation}
\langle a^{P*}_{\ell m} a^{P'}_{\ell' m'}\rangle \equiv \hat{C}^{PP'}(\ell)=
\left(C^{PP'}(\ell)+ N^P(\ell) \right)\delta_{\ell\ell'}\delta_{mm'}.
\end{equation}
Due to statistical isotropy and homogeneity the spectra are diagonal in $\ell$ and $m$. 
Stage III CMB experiments such as WMAP {\citep{Hinshaw2013}} or Planck {\citep{Planck2015_XIII}} will eventually be surpassed by stage IV CMB experiments {\citep{CMBS4,Thornton2016}} which will have a very small instrumental noise (cf. \autoref{tab:cmb}) allowing for measurements up to $\ell \sim 5000$ especially for the polarisation maps. {\newkarl{Very small scales with $\ell>2500$ will thus not be dominated by the instrumental noise. However, due to photon diffusion the signal is strongly suppressed at those scales.. We therefore decide to cut all CMB measurements at $\ell \approx 2500$.}}
It should be noted that the primordial $B$-mode signal is vanishing at the last scattering surface in the analysis here. However, there is a non-vanishing $B$-mode signal induced by gravitational lensing.

{\karl{For the reionization we use a simple hyperbolic tangent with one step for hydrogen and a second for helium reionization. For details on the model we refer to \citet{Lewis2008}}.}

\subsection{Cosmic shear}
Bundle of light rays travelling from distant galaxies get distorted at the percent level due to the perturbed gravitational potentials of the large-scale structure \citep[for reviews, we refer to][]{Bartelmann2001, Hoekstra2008, Bartelmann2010}. Since photons move on null geodesics, lensing measures both gauge invariant quantities of the perturbed metric $\Phi$ and $\Psi$ \citep[up to a sign, depending on the signature of the metric,][]{acquaviva_weak_2004}. In case of general relativity both potentials are equal, but this is no longer the case for modified gravity theories, thus making cosmic shear a probe of deviations from general relativity. The weak lensing effect is described by a line-of-sight integral of the two metric perturbations which defines the lensing potential:
\begin{equation}\label{eq:lensingpot}
\psi_i = \int_0^{\chi_H}\mathrm{d}\chi W_{\psi_i}(\chi) (\Phi +\Psi)\;,
\end{equation}
where $\chi_H=c/H$ is the Hubble radius and $i$ is the index of the tomographic bin. Here the lensing weight function is defined via
\begin{equation}
W_{\psi_i}(\chi) = \frac{G_i(\chi)}{a\chi}\;.
\end{equation}
The tomographic bins are introduced to partially regain redshift information which is destroyed by the line-of-sight integration \citep{Hu2002a}. This procedure is for example not necessary in 3D weak lensing \citep{Heavens2003} where the redshift information is completely carried through the analysis by means a spherical Fourier-Bessel decomposition of the shear field \citep[see in particular][for an application to Horndeski gravity]{SpurioMancini2018}. However, 3D weak lensing is computationally much more expensive than 2D cosmic shear with tomography due to necessary integrations over spherical Bessel functions. 

The tomographic lensing efficiency function is given by
\begin{equation}
G_i(\chi) = 
\int _{\mathrm{min}(\chi,\chi_i)}^{\chi_{i+1}}\mathrm{d}\chi'p(\chi')\frac{\mathrm{d}z}{\mathrm{d}\chi'}\left(1-\frac{\chi}{\chi'}\right)\;,
\end{equation}
with the Jacobi determinant $\mathrm{d}z/\mathrm{d}\chi' = H(\chi')/c$ due to the transformation of the redshift distribution $p(z)\mathrm{d}z$ of background galaxies in redshift $z$, which is given by \citep{Laureijs2011}
\begin{equation}
p(z)\mathrm{d}z \propto z^2\exp\left[-\left(\frac{z}{z_0}\right)^\beta\right]\;.
\end{equation}
Typical parameters for stage IV experiments are $z_0\approx 1$ and $\beta = 3/2$. The angular power spectrum of the lensing potential is then given by
\begin{equation}\label{eq:cosmic_shear_power}
C_{\psi_i\psi_j}(\ell) = 
\int_0^{\chi_H} \frac{\mathrm{d}\chi}{\chi^2}W_{\psi_i}(\chi)W_{\psi_j}(\chi)P_{\Phi+\Psi}(\ell'/\chi,\chi)\;.
\end{equation}
Here we defined the power spectrum of the the sum of the two Bardeen potentials $\Phi$ and $\Psi$.

The noise contribution of observed lensing spectra is Poissonian shape noise due to the finite number of galaxies in each bin which are used to estimate the lensing signal
\begin{equation}
\hat{C}_{\psi_i\psi_j} = C_{\psi_i\psi_j} + \sigma_\epsilon^2\frac{n_\mathrm{bin}}{\bar{n}}\delta_{ij},
\end{equation}
with the intrinsic ellipticity dispersion $\sigma_\epsilon = 0.3$, the number of tomographic bins $n_\mathrm{bin}$ and the mean number density of galaxies $\bar{n}$. Note that, using this definition of the noise, the source distribution must be normalised in each tomographic bin separately. Finally, the convergence spectra is related to the lensing potential spectra via the Poisson equation $\Delta\psi = 2\kappa$, yielding a factor $\ell^4/4$ in the flat-sky limit. By combining geometric information with the growth history and the initial fluctuation spectrum, gravitational lensing is an ideal probe for investigating gravity on cosmological scales \citep{vanderveld_testing_2012, pratten_3d_2016, SpurioMancini2018}, and its second order corrections carry sensitivity with respect to changes in the gravitational interaction \citep{vanderveld_second-order_2011, renk_galileon_2017}.

In this work we will ignore the contribution from intrinsic alignments leading to intrinsically correlated shapes of galaxies \citep[e.g.][]{heavens_intrinsic_2000, croft_weak-lensing_2000, catelan_intrinsic_2001, mackey_theoretical_2002, jing_intrinsic_2002, forero-romero_cosmic_2014, joachimi_galaxy_2015, kiessling_galaxy_2015, kirk_galaxy_2015, blazek_separating_2012, blazek_beyond_2017, Tugendhat2017}. There are two {\karl{additional signals on top of the lensing signal, they are called II and GI alignment and describe the correlation of the intrinsic ellipticity with itself and the correlation of the lensing signal with the intrinsic ellipticity, respectively.}} Both quantities are present in observed ellipticity spectra, which would ideally be interpreted as being proportional to the cosmic shear signal only. Intrinsic alignments only make up a few percent of the lensing signal, depending on the angular scale considered. However, due to the high statistical significance with which cosmic shear is measured by stage IV surveys, the impact on the inference process can be very strong. Additionally, the alignment of galaxies is itself a probe of the LSS and thus contains valuable cosmological information if modelled correctly.

\subsection{Galaxy clustering}
Complementary to cosmic shear, galaxy clustering \citep[e.g.][]{Baumgart1991, Feldman1994, Heavens1995, DiDio2016, Raccanelli2016} measures the statistics of the density contrast, $\delta$, and thus directly the matter power spectrum. Therefore galaxy clustering is only affected by the time-time component of the metric perturbations. However, galaxies are a biased tracer of the total matter distribution \citep[as summarised by][]{desjacques_large-scale_2016}. Quite general we will therefore write $\delta(k,z)b(k,z) = \delta_g(k,z)$. Once the bias is known, we proceed in analogy to the cosmic shear case and find the angular power spectrum of galaxy clustering
\begin{equation}
C_{g_ig_j}(\ell) = \int_0^{\chi_H} \frac{\mathrm{d}\chi}{\chi^2}W_{g_i}(\ell'/\chi,\chi)W_{g_j}(\ell'/\chi,\chi)P_\delta(\ell'/\chi,\chi),
\end{equation}
where $P_\delta$ is the matter power spectrum.
The galaxy weight function is defined as
\begin{equation}
W_{g_i}(\ell/\chi,\chi) = 
    \frac{H(\chi)}{c}b(\ell/\chi,\chi)p(\chi) \  \mathrm{if} \ \chi\in [\chi_i,\chi_{i+1}).
\end{equation}
Thus we assume no cross correlation between different tomographic bins in a galaxy survey. {\karl{This assumption relies heavily on the error of the redshift estimation, which can be quite substantial for photometric surveys. A commonly used value is $\sigma(z) = 0.05(1+z)$. In contrast to cosmic shear, galaxy clustering has a much shorter correlation length and cross-correlations between the bins become only important if the average bin width is of the same order as the redshift error. Here we choose the redshift bins such that an equal number of galaxies lies into each bin and thus the signal-to-noise stays comparable. For the six tomographic bins used in this case this amounts to a minimum redshift width of $\Delta z =0.2$ at $z=1$, showing that only correlation between the two bins at the peak of the redshift distribution will play a significant role. Here we will neglect this contributions, but we keep in mind that this cross-correlation needs to be taken into account for higher $n_\mathrm{bin}$.}} \citet{Bailoni2017} the effect of redshift bin uncertainty and bin cross-correlation was investigated {\karl{for a spectroscopic survey}}. Showing that the impact on the Fisher matrix can be up to $30\%$. 

For the galaxy bias we assume a simple linear model with no scale dependence \citep{Ferraro2015}:
\begin{equation}
b(\chi) = b_0(1+z(\chi))\; .
\end{equation}
{\karl{This model is of course really optimistic, since it does not allow for a non-linear redshift evolution of the bias and has no scale dependence. Since the overall bias amplitude is a free parameter we expect that the influence on $\Omega_\mathrm{m}$ and $\sigma_8$ due to more complex bias models is marginal. It should, however, be noted that the fixed redshift dependence can lead to underestimation of the errors of parameters which depend strongly on the redshift information in the survey, for example $w_0$ and $w_\mathrm{a}$.}}
The observed spectrum suffers again from Poissonian shot noise:
\begin{equation}
\hat{C}_{g_ig_j} = C_{g_ig_j} + \frac{n_\mathrm{bin}}{\bar{n}}\delta_{ij}.
\end{equation}
{\karl{Despite the effects mentioned, there exist further effects such as redshift-space distortion or lensing magnification \citep{Yoo2009}. In our analysis we neglect those effects, which would also influence the modelling of the intrinsic alignment signal.}}

\subsection{CMB lensing}\label{sec:CMBlens}
The CMB lensing signal is induced by the LSS \citep[e.g.][]{Hirata2003,Lewis2006}. Being released at a redshift of roughly $z_*\approx 1100$ the lensing signal of the CMB carries a lot of cosmological information. CMB lensing effectively shuffles patches of the CMB map around differentially, thus destroying the homogeneity of the unlensed CMB. 
Since the unlensed CMB is not accessible, one has to assume that it is indeed statistically homogeneous. In this way can be reconstructed from the five observed spectra (the $EB$ and $TB$ cross-correlations vanish due to parity). A minimal variance and unbiased estimator was constructed by \citet{Hu2002,Okamoto2003}
\begin{equation}
\hat{C}_{DD}(\ell) = {C}_{DD}(\ell) + N_{DD}(\ell)\; .
\end{equation}
Even with future experiments the noise will start dominating at multipoles well below 1000. Thus, the main contribution to the lensing signal comes from linearly evolved structures.
{\karl{The noise properties of the lensing signal are
completely determined by the reconstruction if the lesning signal and its expression is also given in \citet{Hu2002,Okamoto2003}. For the reconstruction noise we use the noise properties of the other estimators as listed in \autoref{tab:cmb}. Furthermore there exists in principle a non-vanishing cross-correlation between the reconstructed lensing signal and the CMB polarization which is not considered here. The largest effect arises from the $B$-mode signal of the CMB polarizations which are then usually discarded to circumvent the problem of dealing with the complicated covariances. In this analysis we ignore this effect since it wont change our analysis strongly as most of the CMB polarization signal arises from the $E$-modes.}}

Let $\chi_*$ be the comoving distance to the last scattering surface, the lensing signal of the CMB is just
\begin{equation}
\psi_\mathrm{CMB} = \int_0^{\chi_H} \mathrm{d}\chi W_\mathrm{CMB}(\Phi +\Psi), 
\end{equation}
with the CMB lensing efficiency function
\begin{equation}\label{eq:CMB_lens_weight}
W_\mathrm{CMB}(\chi) = \frac{\chi_* - \chi}{\chi_*\chi}\frac{H(\chi)}{ca}. 
\end{equation}
The angular power spectrum for the CMB lensing signal has the same structure as Eq. (\ref{eq:cosmic_shear_power}) with the weight function replaced by (\ref{eq:CMB_lens_weight}).

\subsection{Cross-correlations}
Tn the very end, the statistics of the random field describing the LSS is probed, in particular at very different epochs, making the combination of the different observations so powerful. At the same time there are numerous cross-correlations between the different probes. The potential landscape at the last scattering surface introduces correlations between the temperature and polarisation anisotropies. Only the $E$ and $B$ modes are uncorrelated due to parity arguments. 

The calculation of the cross-spectra follows the same recipe as before, with weight functions and power spectrum adopted accordingly. As an example, consider the cross-correlation between cosmic shear and galaxy clustering:
\begin{equation}\label{eq:galshear}
C_{g_i\psi_j}(\ell) = 
\int_0^{\chi_H}\frac{\mathrm{d}\chi}{\chi^2}W_{g_i}(\ell'/\chi,\chi)W_{\psi_j}(\ell'/\chi)P_{\delta,\Phi+\Psi}(\ell'/\chi,\chi).
\end{equation}
Here we defined the cross correlation power spectrum of the density fluctuations with the sum over the two Bardeen potentials. As in the case for cosmic shear we could also rewrite Eq. (\ref{eq:galshear}) so that it only contains $P_\delta$ by using the ratio between $\Psi$ and $\Phi$, as well as the modified Poisson equation. 

{\newkarl{For the noise part we consider cells $i$ with population number $n_i$ and the quantity}}
\begin{equation}
f(\boldsymbol{x}) = \sum _i \delta_\mathrm{D}(\boldsymbol{x}-\boldsymbol{x_i})n_if_i\; ,
\end{equation}
{\newkarl{which can be cast into galaxy counts, $\delta_{g_i}$, or a shear measurement, $\epsilon_i$, respectively. Working out the correlation $\langle f f\rangle$ one finds, as expected, that the shot-noise is sourced by the correlation $\langle\epsilon_i \delta_{g_i}\rangle$. This correlation in principle should vanish, since the intrinsic shape is not correlated with the galaxies position and the average number density. However, considering intrinsic alignment models for linear galaxies, the intrinsic shape is sourced by the gravitational tidal field \citep{troxel_intrinsic_2015}. Therefore the intrinsic ellipticity can be related to the density. Consequently there will be some additional correlation between galaxy clustering and weak lensing due to the intrinsic alignment model. However, we expect this contribution to be very small compared to the overall signal.}}

Additionally, galaxy clustering as well as cosmic shear are correlated with the CMB lensing signal \citep{kitching_3d_2014}. Lensing of the CMB and the temperature anisotropies have a non-vanishing cross-correlation due to the integrated Sachs-Wolfe effect \citep{lewis_weak_2006}.

On very large scales the temperature fluctuations of the CMB are modified due to time-evolving potentials. This is called the integrated-Sachs-Wolfe effect {\citep[iSW,][]{Sachs1967}}. 
The weight function of the iSW effect is given by
\begin{equation}
W_\mathrm{iSW}(k,a) = \frac{3}{2\chi_H^3}a^2E(a)F'(k,a),
\end{equation}
where the prime denotes a derivative with respect to $a$ and
\begin{equation}
F(k,a) = \mu(k,a)\frac{D_+(k,a)}{a}\left(1+\frac{1}{\gamma(k,a)}\right).
\end{equation}
Since the ISW is a late-time effect there is a cross-correlation between galaxy clustering, cosmic shear and CMB lensing. For the first we obtain the following cross-spectrum as
\begin{equation}
C_{g_iT}(\ell) = \int \frac{\mathrm{d}\chi}{\chi^2}W_{g_i}W_\mathrm{iSW} D_+(\ell'/\chi,\chi)\frac{P_\delta(\ell'/\chi)}{k^2} 
\end{equation}
and similarly for $C_{\psi_i T}$ and $C_{DT}$.

\begin{figure*}
\begin{center}
\includegraphics[width = .45\textwidth]{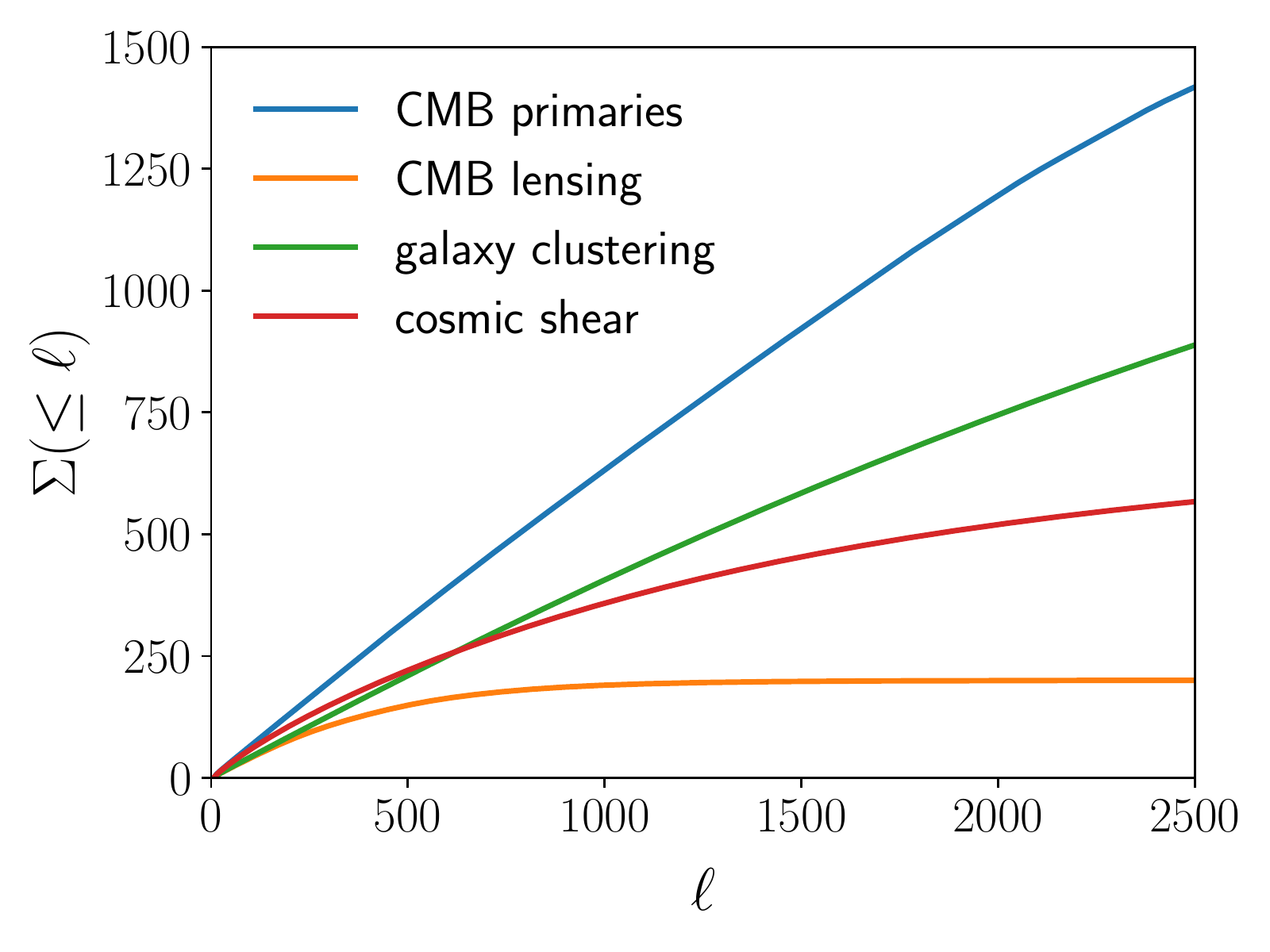}
\includegraphics[width = .45\textwidth]{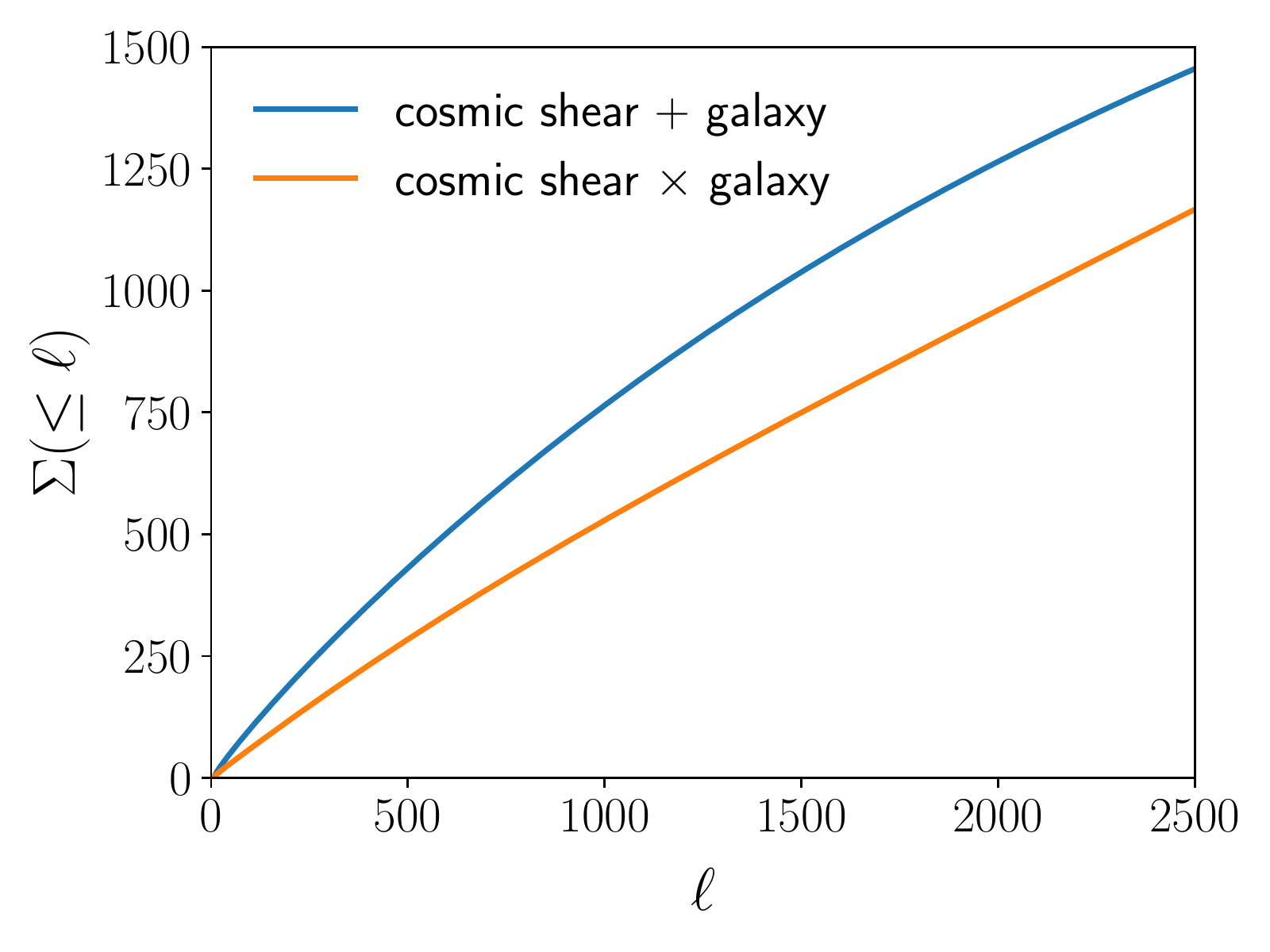}
\caption{Cumulative signal-to-noise ratio (see Eq.~\ref{eq:snr}). For a survey with settings as in \autoref{table:experimental_setup} but with $n_\mathrm{bin} = 1$. \textit{Left}: All probes individually. \textit{Right}: Adding galaxy clustering and cosmic shear with ($\times$) and without ($+$) cross-correlation.}
\label{fig:StoR}
\end{center}
\end{figure*}

\section{Statistics}\label{sec:statistics}
We collect all probes in a complex vector $\boldsymbol{a}_{\ell m}$, such that the statistics of the modes can be represented by the covariance matrix, $\boldsymbol{C}_{\boldsymbol{a}}(\ell) \equiv \langle \boldsymbol{a}_{\ell m} \boldsymbol{a}_{\ell m}^+\rangle$.
The likelihood of a set of modes to be represented by its covariance is given by
\begin{equation}\label{eq:likeli_gauss}
\mathcal L (\{\boldsymbol{a}_{\ell m}\}) = \prod_\ell p\left[\left\{\boldsymbol{a}_{\ell m}\right\}|\boldsymbol{C}_{\boldsymbol{a}}(\ell)\right]^{2\ell+1}\; .
\end{equation}
Due to statistical isotropy and homogeneity the likelihood decomposes into disconnected products of $\ell$- and $m$-modes. The exponent, $2\ell+1$, takes care of the multiplicity of the $m$-modes as a reflection of statistical isotropy. Note that the covariance implicitly depends on the model parameters.

\begin{table}
\begin{center}
\begin{tabular}{c|c|c|c|c|c}
 $n_0$ & $\bar{z}$ & $n_\mathrm{bin}$ & $f_\mathrm{sky}$ & $\ell_\mathrm{min}$ & $\ell_\mathrm{max}$ \\
\hline
30$\,\mathrm{arcmin}^{-2}$ & $0.9$ & $6$ & $15.000\,\mathrm{deg}^2$ &$10$ & $2000$, $2500$ 
\end{tabular}
\caption{Basic parameters of the experimental setup for a stage IV experiment. While the noise properties of galaxy clustering and cosmic shear is controlled by $n_0$, $n_\mathrm{bin}$ and $\sigma_\epsilon \approx 0.3$, the noise of CMB anisotropies depends on various parameters, which are summarised in \citet{Thornton2016}. {\newkarl{The two values for $\ell_\mathrm{max}$ refer to LSS and CMB observations respectively.}}}
\label{table:experimental_setup}
\end{center}
\end{table}

Constructing the likelihood therefore reduces to a model for the distribution of the modes. At early times, the modes can be assumed to be Gaussian to very good approximation. On small scales at late times, however, Gaussianity is destroyed due to non-linear evolution. Empirically, the modes seem to follow a log-normal distribution \citep{Hilbert2011}, but to date there is no analytic derivation of this behaviour from first principles. A way out is to not consider the modes, but rather to compress the data to obtain an estimator of the power spectrum, $\hat{C}(\ell_i) = \sum_{\ell\in\ell_i}|a_{\ell m}|^2$, where no information is lost due to statistical isotropy. The advantage of this estimator is that the distribution is very well approximated by a Gaussian at sufficiently large $\ell$ due to the central limit theorem. {\karl{At small multipoles the distribution is skewed and can be shown to follow a Gamma distribution if the modes themselves are distributed in a Gaussian way (\citet{Hamimeche2009} for CMB likelihoods and \citet{Sellentin2017,Sellentin2017b} for cosmic shear likelihoods)}}.
Another advantage of the latter approach is that the influence of probes can be analysed more independently. At the same time, the construction of the likelihood for the power spectrum estimator requires the knowledge of its covariance, which itself depends on the cosmological parameters \citep[e.g.][]{Scoccimarro1999}. This makes the inference process more complicated, since higher order cumulants have to be calculated.

\begin{table}
\begin{center}
\begin{tabular}{c|c|c}
$\nu[\mathrm{GHz}]$ & $\theta_\mathrm{beam}[\mathrm{arcmin]}$ & $\sigma_P[\mu\mathrm{K}\;\mathrm{arcmin}]$ \\
\hline
90 & 5.7 & 18.80 \\
105 & 4.8 & 13.80 \\
135 & 3.8 & 9.85 \\
160 & 3.2 & 7.78 \\
185 & 2.8 & 7.05 \\
200 & 2.5 & 6.48 \\
220 & 2.3 & 6.26 \\
\end{tabular}
\end{center}
\caption{Noise level for a stage IV CMB experiment in different bands. The total noise contribution is then the inverse weighted sum of the individual noise contribution in each band.}
\label{tab:cmb}
\end{table}

Here, we will assume the modes to follow a Gaussian distribution for simplicity, and that the statistical symmetries of homogeneity and isotropy are fulfilled. This implies no correlation between $\ell$- or $m$-modes, and we correct the covariance for an incomplete sky coverage with a factor $\sqrt{f_\mathrm{sky}}$. Consequently, the distribution of a single mode in Eq.~(\ref{eq:likeli_gauss}) is given by
\begin{equation}
p\left(\left\{\boldsymbol{a}_{\ell m}\right\}|\boldsymbol{C}_{\boldsymbol{a}}(\ell)\right) = \frac{1}{\sqrt{(2\pi)^N\mathrm{det}\boldsymbol{C}(\ell)}}\exp\left[-\frac{1}{2}\boldsymbol{a}^+_{\ell m}\boldsymbol{C}_{\boldsymbol{a}}^{-1}(\ell)\boldsymbol{a}_{\ell m}\right]\; .
\end{equation}
Here, $N = 4 + 2n_\mathrm{bin}$ is the dimension of the data vector $\boldsymbol{a}_{\ell m}$. 

Having said this it is already clear that by combining different measurements and thus a new entry to $\boldsymbol{a}_{\ell m}$ all possible correlations with the other measurements must be explained by the model. 
This effect is subtle: the non-vanishing correlations lead to a decreasing signal, compared to measurements which are independent. However, as said before, the model has to explain those correlations as well, possibly yielding a larger constraining power. It is thus necessary to include all correlations, since ignoring them would imply the wrong assumption of statistical independence and of vanishing cross-correlations which would be predicted to be nonzero by a given cosmological model.

The logarithmic likelihood $L = -2\mathrm{ln} \mathcal{L}$ can be brought into the following form
\begin{equation}
L = \sum_\ell (2\ell +1)\left[\text{ln}\left(\text{det}\boldsymbol{C}_{\boldsymbol{a}}(\ell)\right) + \boldsymbol{a}^+_{\ell m} \boldsymbol{C}_{\boldsymbol{a}}^{-1}(\ell)\boldsymbol{a}_{\ell m}\right]\; , 
\end{equation}
up to an irrelevant additive constant. Averaging over the data yields
\begin{equation}\label{eq:loglike}
\langle L\rangle = \sum_\ell (2\ell+1)\left[\text{ln}\left(\text{det}\boldsymbol{C}_{\boldsymbol{a}}\right) + \text{tr}(\boldsymbol{C}_{\boldsymbol{a}}^{-1}\boldsymbol{\hat C}_{\boldsymbol{a}})\right]\; ,
\end{equation}
where $\boldsymbol{\hat C}_{\boldsymbol{a}} = \langle \boldsymbol{a}_{\ell m}\boldsymbol{a}^+_{\ell m}\rangle$ is the covariance matrix of the data. In our case the covariance assumes the following form for each $\ell$ separately and independent of $m$, $\boldsymbol{\hat C}_{\boldsymbol{a}}=$
\begin{equation*}
\begin{pmatrix}
TT & TE & 0 & TD & Tg_1 & \dots & Tg_{n}  & T\gamma_{1} &  \dots & T\gamma_n \\ 
ET & EE & 0 & ED & 0 & \dots & 0  & 0 &  \dots & 0 \\
0 & 0 & BB & 0 & 0 & \dots & 0  & 0 &  \dots & 0 \\  
DT & DE & 0 & DD & Dg_{1} & \dots & Dg_{n}  & D\gamma_{1} &  \dots & D\gamma_n \\
g_1T & 0 & 0 & g_1D & g_1g_1 & \dots & g_1g_{n}  & g_1\gamma_{1} &  \dots & g_1\gamma_n \\  
\vdots & \vdots & \vdots & \vdots & \vdots & \ddots & \dots & \vdots & \ddots & \vdots \\
g_nT & 0 & 0 & g_nD & g_ng_n & \dots & g_ng_{1}  & g_n\gamma_{n} &  \dots & g_n\gamma_n \\  
\gamma_1T & 0 & 0 & \gamma_1D & \gamma_1 g_1 & \dots & \gamma_1g_{n}  & \gamma_1\gamma_{1} &  \dots & \gamma_1\gamma_n \\  
\vdots & \vdots & \vdots & \vdots & \vdots & \ddots & \dots & \vdots & \ddots & \vdots \\
\gamma_nT & 0 & 0 & \gamma_nD & \gamma_n g_1 & \dots & \gamma_ng_{n}  & \gamma_n\gamma_{1} &  \dots & \gamma_n\gamma_n 
\end{pmatrix}.
\end{equation*}
where $n$ is the number of tomographic bins and the label of each entry $XY$ denotes the covariance $C^{XY}(\ell)$ discussed in section \ref{sec:data}. The symmetry and positive definiteness of the covariance is ensured by the Cauchy-Schwarz inequality. It should be noted that there is no correlation $\langle\gamma_i g_j\rangle$, for $z_i < z_j$. 
 
\begin{figure}
\begin{center}
\includegraphics[width = .49\textwidth]{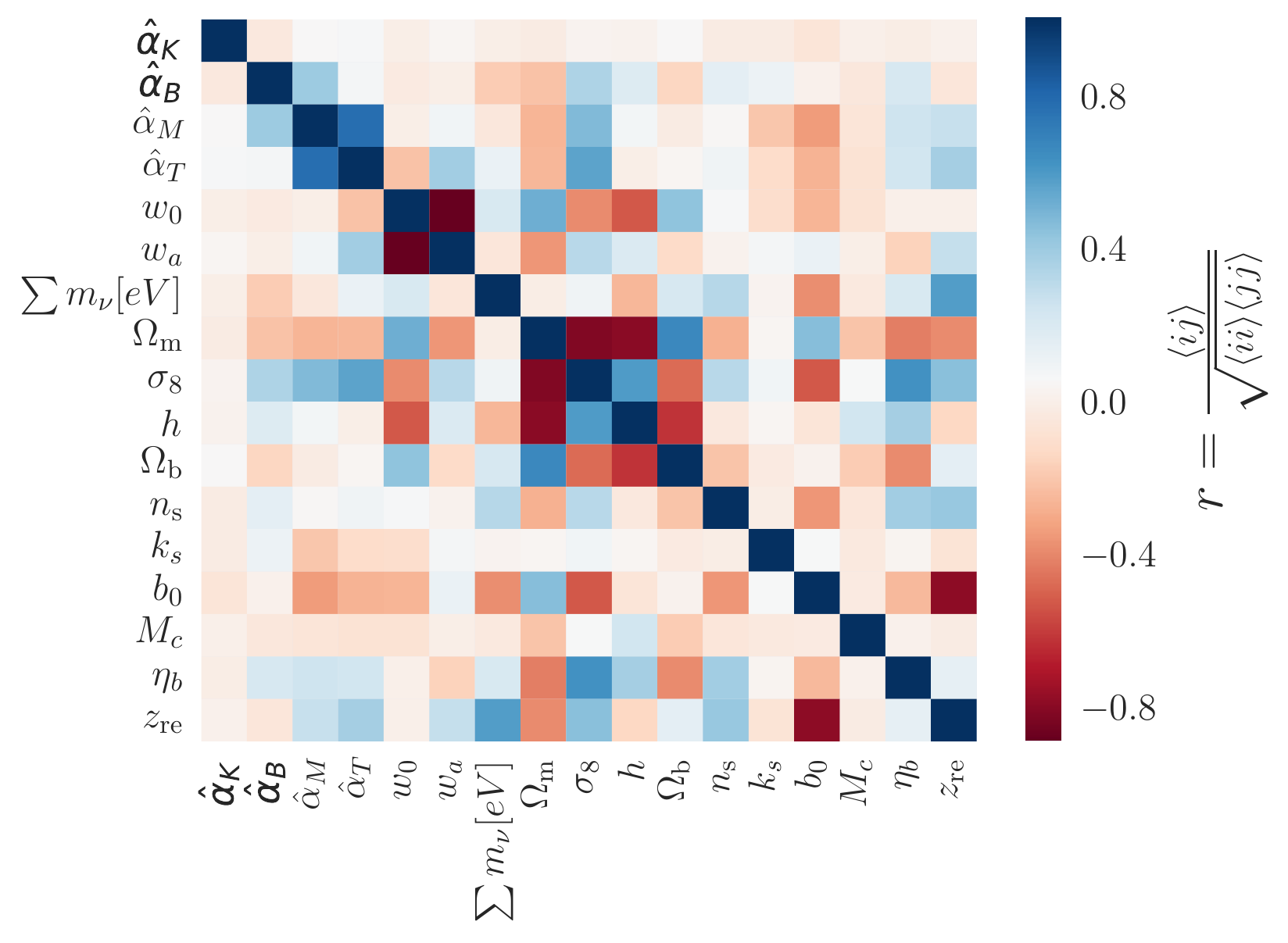}
\caption{Pearson correlation coefficient $r_{\mu\nu}$ for all parameters in the analysis (including $\hat{\alpha}_\mathrm{K}$ and $\hat{\alpha}_\mathrm{T}$) for a survey with specifications as in \autoref{table:experimental_setup}.}
\label{fig:matrix_correlation}
\end{center}
\end{figure} 
 
The posterior equals the likelihood for a uniform prior up to a normalization constant by virtue of Bayes' theorem and the maximum-likelihood estimate is defined to be the point in parameter space $\hat{\boldsymbol{\theta}}$, where $\langle L\rangle$ has its maximum.
 
A common approximation for the posterior is the Fisher-matrix $F_{\mu\nu}$, which is the curvature of the log-likelihood at the best fit value. For Gaussian distributed data it can be calculated easily by \citep{Tegmark1997}:
\begin{equation}\label{eq:fisher}
F_{\mu\nu} = 
\sum_{\ell = \ell_\mathrm{min}}^{\ell_\mathrm{max}}
\frac{2\ell +1}{2}\mathrm{tr}\left(\partial_\mu \mathrm{ln} [\boldsymbol{C}_{\boldsymbol{a}}(\ell)]\:\partial_\nu \mathrm{ln} [\boldsymbol{C}_{\boldsymbol{a}}(\ell)]\right)\; .
\end{equation}
The Fisher-matrix is exact only if the posterior distribution is Gaussian. It is especially insufficient for poorly constrained parameters and high-dimensional parameter spaces. The Cram\'{e}r-Rao inequality states that the Fisher-matrix yields a lower bound for the forecasted errors only. Furthermore, the Fisher-matrix can be used to calculate the cumulative signal-to-noise ratio:
\begin{equation}\label{eq:snr}
\Sigma^2(\leq \ell) = 
\sum_{\ell = \ell_\mathrm{min}}^{\ell_\mathrm{max}} 
\frac{2\ell+1}{2}\mathrm{tr}\left(\boldsymbol{C}^{-1}_\mathrm{a}(\ell)\boldsymbol{S}_\mathrm{a}(\ell)\boldsymbol{C}^{-1}_\mathrm{a}(\ell)\boldsymbol{S}_\mathrm{a}(\ell)\right)\; ,
\end{equation}
where $\boldsymbol{S_a}$ is the covariance without noise. Effectively, the signal to noise-ratio $\Sigma$ corresponds to the ratio between the signal of an assumed amplitude $A$ divided by the anticipated statistical error $\sigma_A = 1/\sqrt{F_{AA}}$ from the Fisher-matrix $F_{AA}$. {\karl{The derivative in Eq. (\ref{eq:fisher}) is calculated by finite differencing:}}
\begin{equation}
\partial_\mu f(x^\mu) \approx \frac{f(x^\mu +\Delta x^\mu)-f(x^\mu -\Delta x^\mu)}{2\Delta x^\mu}\; .
\end{equation}
{\karl{Convergence is ensured by varying $\Delta x^\mu$ until $F_{\mu\nu}$ does not change by more than $5\%$.}}
All expressions so far assumed that the spherical harmonic coefficients completely describe the fields under consideration. This is true only if the corresponding basis functions form a complete basis. If the survey area does not cover the whole sky this is no longer true, and especially for small survey areas and complicated masks this effect can become very important. For state IV surveys we expect the sky-coverage $f_\mathrm{sky}$ to be roughly one third. It therefore suffices to multiply the log-likelihood with a factor of $f_\mathrm{sky}$ to model the information loss.

Additionally to the Fisher analysis we carry out a MCMC using the sampling technique proposed by \citet{Goodman2010}. This stretch-move technique is invariant under affine transformation of the parameter space and is rather insensitive on the covariance between the parameters. {\karl{Convergence of the chains is ensured by estimating the exponential auto-correlation time and letting the walkers run for multiple auto-correlation times after the initial burn-in period.}}
For a more detailed discussion we refer to \citet{Redlich2014}.

\section{Constraints on modified gravity}\label{sec:results}
\begin{figure}
\begin{center}
\includegraphics[width = .45\textwidth]{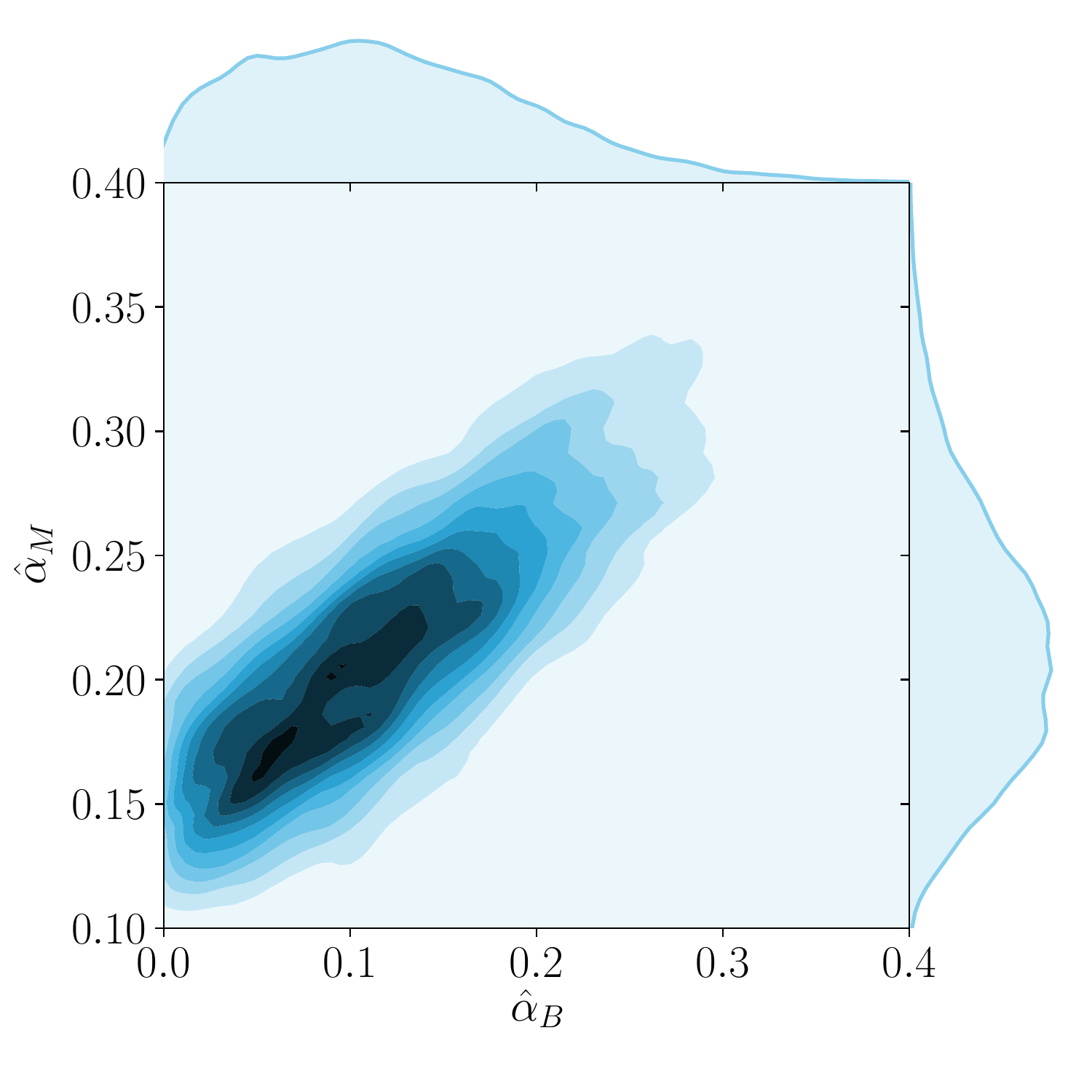}
\caption{Joint plot of the resulting posterior estimated from the MCMC for the remaining free modified gravity parameters. All other parameters have been marginalised over. Note that the contours shown do not correspond to certain $\sigma$-regions.}
\label{fig:amabmcmc}
\end{center}
\end{figure}

Likelihoods of the parameter set of are evaluated using MCMC-sampling. In particular we employ an affine-invariant sampler. The results are then compared to the Fisher matrix forecast, which effectively assumes a Gaussian posterior. The latter comparison is important because numerical errors in the evaluation of the Fisher matrix might lead to differing parameter degeneracies and because the Fisher matrix technique only allows for lower bounds on the magnitude of statistical errors.\autoref{fig:amabmcmc} shows joint constraints on the two Horndeski parameters $\alpha_M$ and $\alpha_B$, as evaluated by MCMC-sampling, with a marginalisation over the remaining parameters. Clearly, $\alpha_M$ can be measured to be nonzero in a significant way while a measurement of $\alpha_B$ is possible but less significant. For the purpose of this invesigation, we displace the reference cosmology away from the $\Lambda$CDM-parameter choices, but will comment on the relevance of the reference cosmology further below.

\begin{figure*}
\begin{center}
\includegraphics[width = .9\textwidth]{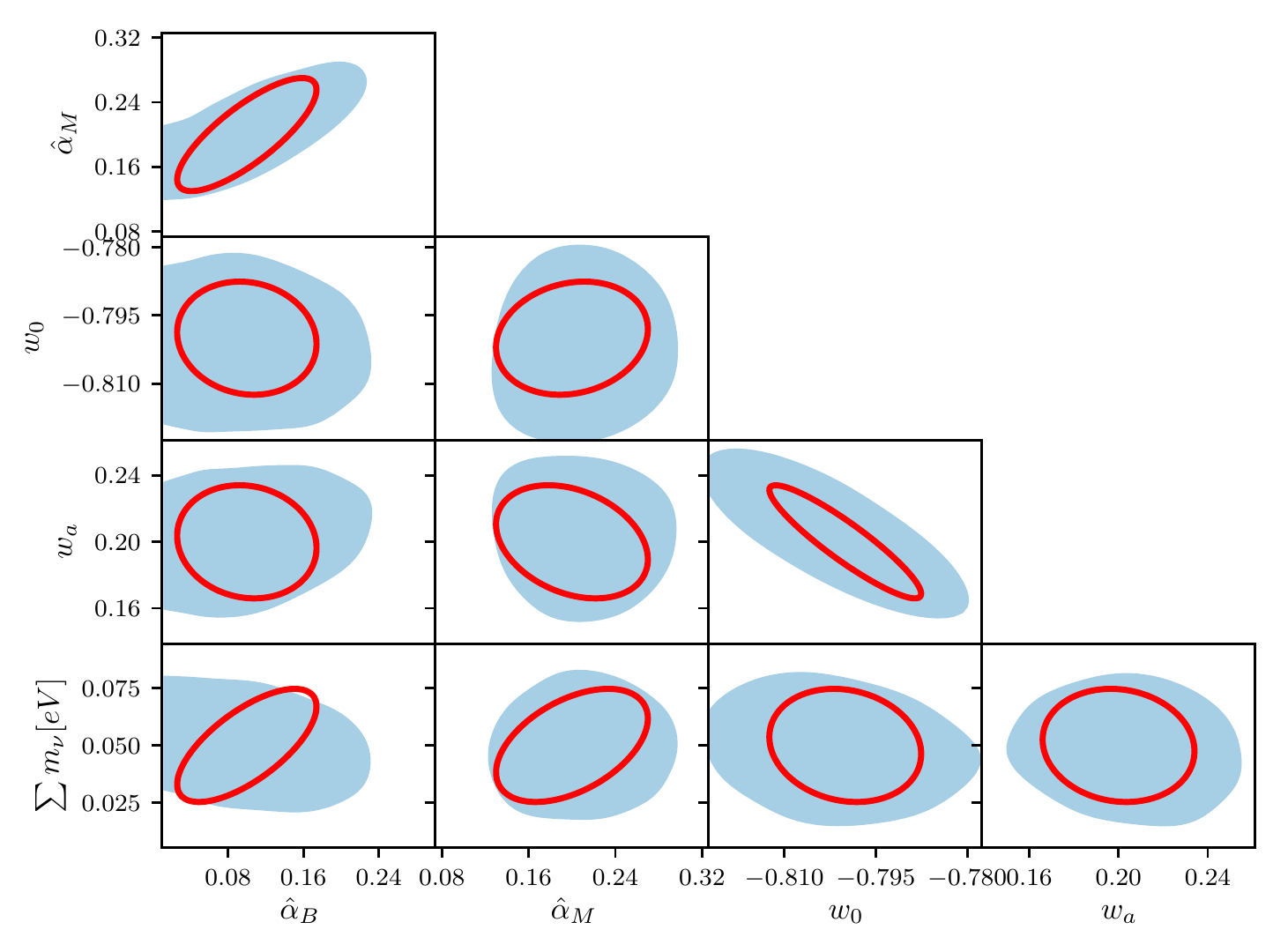}
\caption{Comparison of the MCMC results (light blue) with the Fisher-matrix forecast (red ellipse). The contour plot shows the probability only in the $68\%$ region and can therefore be seen as a direct comparison with the Fisher-matrix.
All contours are marginalised over all other parameters summarised in \autoref{tab:params} where also the marginal errors are given. The experimental setup is the one described in \autoref{table:experimental_setup}.}
\label{fig:compare_Fish_MCMC}
\end{center}
\end{figure*}

\autoref{fig:compare_Fish_MCMC} shows the true MCMC-evaluated likelihood on the two $\alpha$-parameters, the neutrino masses and the dark energy parameterisation in comparison to the results for a Gaussian approximation of the likelihood. We selected these parameters because they are the most difficult to measure with the largest statistical errors and at the same time the most interesting: while any parameter that enters the model in a non-linear way, will have a non-Gaussian likelihood, this non-Gaussianity is more prominent in less constrained parameters than in better constrained ones. While in most cases with a parameter degeneracy the Fisher-matrix formalism seems to be able to predict the reasonably well, the degeneracy between the neutrino mass $\sum m_\nu$ and $\alpha_B$ is not reproduced. Apart from those results, the most important consequence of non-Gaussian likelihoods is that they give rise to larger statistical errors as would be expected from the Cram{\'e}r-Rao bound.

\subsection{Experimental setup}
The experimental characteristics anticipated for stage IV-observations of the cosmic large-scale structure are summarised in \autoref{table:experimental_setup}. They correspond to the data set retrieved by the European Euclid mission \citep{Laureijs2011} and in conjunction with high-resolution observations of the cosmic microwave background of the next generation of CMB observatories one can access multipoles up to $\ell\simeq3000$ in temperature and $\ell\simeq 5000$ in polarization before being limited by noise. {\karl{The noise levels of a stage IV CMB experiment are listed in \autoref{tab:cmb}. It should, however, be noticed that we use a maximum multipole of 2500, since higher multipoles will not contribute strongly to the overall signal-to-noise due to the damping of the CMB anisotropies.}} {\newkarl{This will also have negligible influence on the overall parameter constraints on modified gravity since the sensitivity on gravity originates from larger scales.}}

{\karl{
As a lower multipole limit we use $\ell_\mathrm{min} =10$, which is a realistic value for LSS measurements. For CMB measurements the noise curve will not stay flat at very low $\ell$ due to atmospheric effects even for stage IV experiments. This effect is not included in our analysis and, so that the noise level is entirely dominated by cosmic variance only also on very large scales. The main effect of this will be the further limitation of the measurement of the sum of the neutrino masses due to the degeneracy between the fluctuation amplitude and the optical depth.}}

Data of that quality show an impressive overall signal to noise ratio, as depicted by \autoref{fig:StoR} and allow for the determination of the parameter values of a complex cosmological model. The cumulative signal is given for an all-sky observation with no redshift information, i.e. $n_\mathrm{bin}=1$, and adding tomographic information would increase the sensitivity of the measurement with respect to line of sight-varying quantities. Furthermore, cosmic shear is levelling out earlier than galaxy clustering because of the line of sight-integrating nature of weak lensing. CMB anisotropies consist of three probes, namely temperature and $E$-mode polarisation spectra and cross-spectra, with the additional reconstructed CMB-lensing deflection field, for which we assume information originating from the polarisation $B$-mode as well. On large angular scales, all probes are cosmic variance dominated, and the comparatively small signal strength of the CMB-lensing deflection field is caused by the reconstruction noise.

Large scale structure data, i.e. weak gravitational lensing and galaxy clustering, probe overlapping sections of the cosmic large-scale structure and are therefore not statistically independent. The right panel in \autoref{fig:StoR} shows the influence of the non-vanishing cross-correlation between the different probes: Ignoring the cross-correlation would overestimate the signal to noise ratio, as shown by the blue curve, which is corrected by the inclusion of the correct cross-correlation, as shown by the orange line. This reduction in signal strength would amount to $\simeq 16\%$, but would not straightforwardly imply a similar decrease in sensitivity. In contrast, because the cross-correlation is in fact measurable, it constitutes a valuable source of cosmological information.

Even though we find that the signal to noise-ratio still increases for $\ell >2500$, we ignore those scales from our analysis since they probe the deeply non-linear regime, which would require a faithful description of non-linear structure formation, in particular trispectrum corrections to the covariance matrix. Furthermore, we do not consider a cut-off multipole as it is for example done by \citet{Alonso2017} for galaxy clustering.

\begin{table*}
\label{tab:params}
\begin{center}
  \begin{tabular}{c|c|c|c|c|c|c|c}
  \hline
  Parameter &  fiducial (MG) & fiducial ($\Lambda$CDM)  & $\sigma$ (MG) & $\sigma_\mathrm{F}$ (MG) & $\sigma_\mathrm{F}$ ($\Lambda$CDM) &prior &comments \\
  \hline
  \hline 
$\hat\alpha_\mathrm{K}$ & 0.05 & 0.05 & - & - & - & $[0,\infty)$ & fixed \\
$\hat\alpha_\mathrm{B}$ & 0.1 & 0.01 & 0.065 & 0.035 & 0.041 & $[0,\infty)$ & \\ 
$\hat\alpha_\mathrm{M}$ & 0.2 & 0.01 & 0.049 & 0.032 & 0.043 & $[0,\infty)$ & \\
$\hat\alpha_\mathrm{T}$ & 0.01 & 0.01 & -& - & -  & $[0,\infty)$ & fixed\\
$ w_0$ & -0.8 & -1.0 & 0.0134 & 0.006 & 0.009 & $[-1,1]$ & \\
$ w_\mathrm{a}$ & 0.2 & 0.0 & 0.029 & 0.016 & 0.028& $[0,\infty)$ &\\
$ \sum m_\nu$[eV] & 0.05 & 0.05 & 0.016 & 0.011 & 0.01 & $[0,\infty)$ &\\
$\Omega_{\mathrm{m0}}$ & 0.314 &0.314 & 0.001 & 0.0007 & 0.001 & $[0,\infty)$ & \\
$\sigma_8$ & 0.834 & 0.834 & 0.002 & 0.0016 & 0.0013 & $[0,\infty)$ &\\
$h$ & 0.674 & 0.674 & 0.0013 & 0.0008 & 0.009 & $[0,\infty)$  &\\
$\Omega_\mathrm{b}$ & 0.0486 & 0.0486 & 0.00019 & 0.00013 & 0.00016 &$[0,\infty)$  &\\
$n_\mathrm{s}$ & 0.962 & 0.962 & 0.0014 & 0.0013 & 0.0013 &$[0,\infty)$  &\\
$z_\mathrm{re}$ & 11.357 & 11.357 & 0.28 & 0.026 & 0.027 &$[0,\infty)$  & \\
\hline
$k_\mathrm{s}$ & 0.1 & 0.1 & 0.010 & 0.012 & &$[0,\infty)$  &\\
$b$ & 0.68&  0.68 & 0.0016 & 0.0010 & 0.001 &$[0,\infty)$  &\\
$M_\mathrm{c}$ & 0.26& 0.26 & 0.0023 & 0.0023 & 0.0022 & $[0,\infty)$  &\\
$\eta_b$ & 0.5 &  0.5 & 0.0063 & 0.006 & 0.006 &$[0,\infty)$  &\\
\hline
\end{tabular}
\end{center}
\caption{Parameters used for the inference process with two different fiducial models, and the $68\%$ marginalised errors from the Fisher analysis, $\sigma_\mathrm{F}$ and the MCMC $\sigma$. For the MCMC, the errors are just given for the modified gravity-model, while the Fisher analysis covers both the modified gravity- and $\Lambda$CDM fiducial. All uncertainties given correspond to the usage of all probes discussed. The fiducial values are taken from \citet{Planck15parameters}, \citet{Schneider2015} and \citet{Ferraro2015}.}
\end{table*}

\subsection{Cosmological constraints}
The full set of four possible $\alpha$-functions that characterise a Horndeski-cosmology can not be constrained by cosmological data. As shown by \citet{Alonso2017}, the kineticity $\alpha_\mathrm{K}$ is basically unconstrained by cosmological observations and uncorrelated with the remaining modified gravity parameters. Furthermore, the tensor speed excess, $\alpha_\mathrm{T}$, has been constrained recently to be very close to zero by observations of the near-coincidence of the electromagnetic and gravitational wave signal of a neutron star merger {\citep{Abbott2017b}}. Consequently, we exclude both parameters from our analysis and only retain $\alpha_M$ and $\alpha_B$, which form together with the standard FLRW-parameters $\Omega_\mathrm{m}$ and $\Omega_b$, a time-varying dark energy component with equation of state parameters $w_0$ and $w_a$, the parameters $n_\mathrm{s}$ that determine the shape of the CDM-spectrum and determine the physical density parameters, and finally phenomenological and astrophysical parameters such as the reionisation redshift $z_\mathrm{re}$, the biasing parameter $b$ and the baryons-to-photons ratio $\eta_b$. Together, these parameters form a 17-dimensional parameter space which we set out to constrain through cosmological data.

Degeneracies and correlations between the parameters are described by Pearson's correlation coefficient $r_{\mu\nu}$ under the assumption of a Gaussian posterior distribution,
\begin{equation}
r_{\mu\nu} = 
\frac{\mathrm{cov}(\theta_\mu,\theta_\nu)}{\sqrt{\mathrm{cov}(\theta_\mu,\theta_\mu)\mathrm{cov}(\theta_\nu,\theta_\nu)}},
\end{equation}
which can be straightforwardly derived from the Fisher-matrix under the assumption of a Gaussian likelihood, where the parameter covariance is equal to the inverse Fisher-matrix. The results are shown in Fig.~\autoref{fig:matrix_correlation}, where one can indeed see that $\alpha_\mathrm{K}$ is not correlated with the remaining parameters.

As a prior we assume a flat prior for all parameters. Due to physical conditions, the parameter space is bounded for some parameters as summarised in \autoref{tab:params} by the requirement that the parameters should be positive, for example, in order to exclude unstable perturbations if $\alpha_i<0$. 

For the choice of the fiducial model we offer two alternatives, an ad-hoc Horndeski model with the parameter choices $\alpha_B=0.1$, $\alpha_M=0.2$ with an evolving dark energy component with $w_0=-0.8$ and $w_a=0.2$ in comparison to a standard $\Lambda$-CDM cosmology ({\karl{note that the fiducial values are not exactly at the $\Lambda$CDM fiducial due to stability reasons}}), where the $\alpha$- and $w$-parameters are not relevant but where one nevertheless can derive error forecasts. All constraints will vary across parameter space \citep{Schaefer2016}, which motivated us to quote errors and correlation coefficients for different choices of the fiducial model. In forecasting, we consider the parameters compiled in \autoref{tab:params}, where we summarise the expected uncertainties, both for the Fisher- and MCMC-evaluated likelihood, the fiducial choices and the resulting true and Gaussian-approximated statistical errors.

In \autoref{fig:amabmcmc} we show the marginalised posterior for the remaining two modified gravity parameters $\alpha_M$ and $\alpha_B$. There is a strong correlation between $\alpha_\mathrm{M}$ and $\alpha_\mathrm{B}$ as also found in \citet{Alonso2017}. Furthermore, one can see the skewness in the distribution, which is caused by the positivity bound on the two parameters. We expect that the distribution to become more symmetric if the maximum likelihood-estimate is further away from the $\Lambda$CDM fiducial model or if the statistical errors of one of the parameters become less by inclusion of more data.

In \autoref{fig:compare_Fish_MCMC} we show the constraints on the modified gravity parameters, the dark energy equation of state and the neutrino masses for the fiducial Horndeski-cosmology. The blue shaded contours correspond to the distribution estimated from the MCMC corresponding to the $1\sigma$ confidence interval. All other parameters from \autoref{tab:params} are marginalised over. Red ellipses show the corresponding error forecast from the Fisher-fmatrix formalism. Clearly, the Gaussian approximation generally leads to tighter constraints as can be expected from the Cram{\'e}r-Rao-inequality. Especially for more poorly constrained parameters the MCMC errors are significantly larger. For the $1\sigma$ errors, we find the errors to be up to 1.8 times larger for $\alpha_\mathrm{B}$ and 1.3 times for $\alpha_\mathrm{m}$. Furthermore, we find that degeneracies in general, for instance between $\alpha_\mathrm{m}$ and $\alpha_\mathrm{B}$, are well represented by the Fisher-matrix formalism and that the largest impact of non-Gaussian distribution concerns the magnitude of the errors where the Fisher-matrix formalism is overly optimistic. 

Quite generally, we find the degeneracies to be very similar when comparing the Fisher-matrix with the MCMC, in particular the strong degeneracy between $w_\mathrm{a}$ and $w_0$ is reproduced very nicely. The only exception is the degeneracy between the sum of the neutrino masses and the braiding-term which shows a stronger degeneracy in the Fisher forecast. Since every contour shown in \autoref{fig:compare_Fish_MCMC} shows a marginalised contour, it is not straightforward to follow where this discrepancy originates from.

\begin{figure*}
\begin{center}
\includegraphics[width = .9\textwidth]{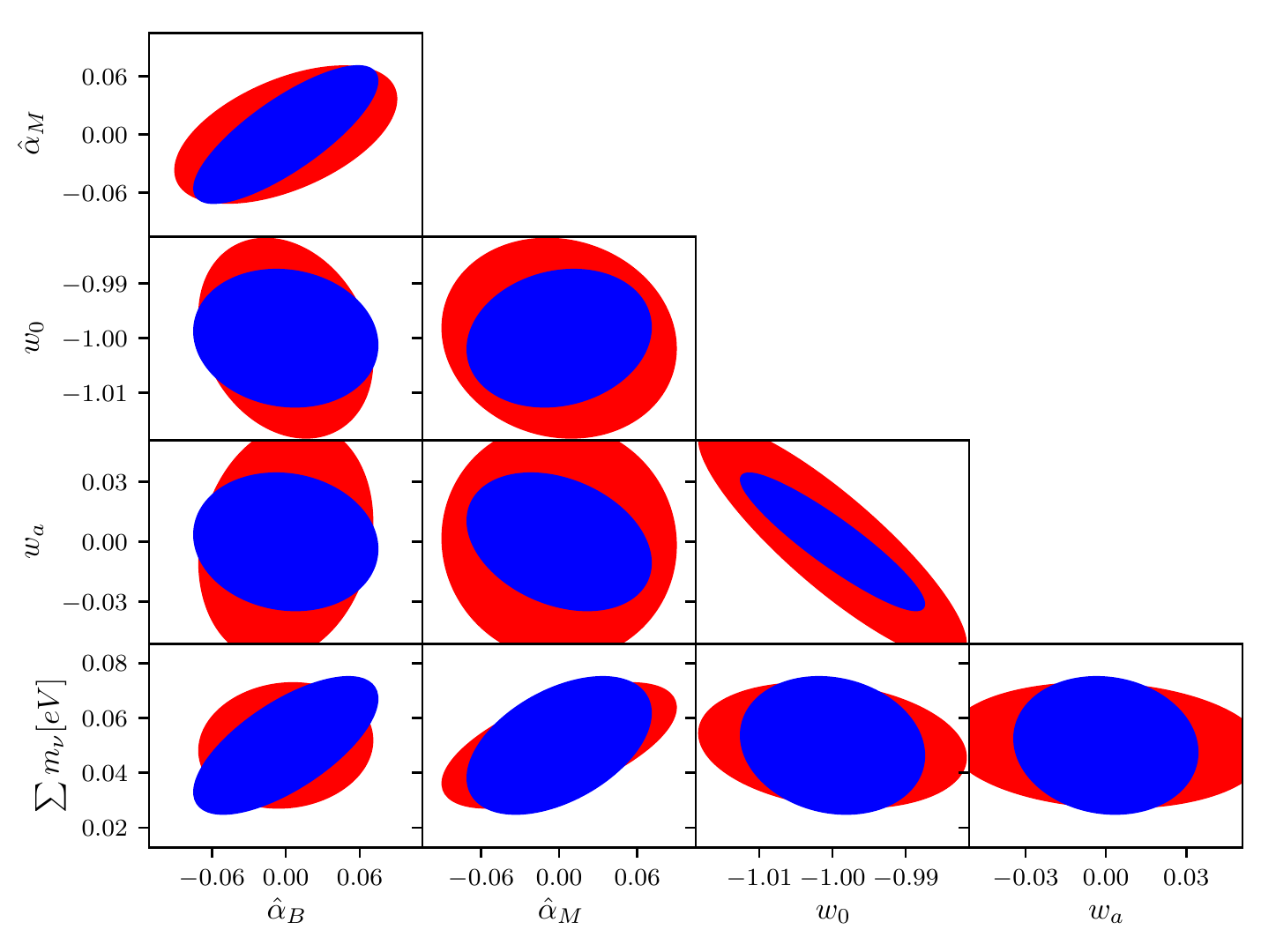}
\caption{Dependence of the Fisher-matrix on the fiducial point in parameter space. Red ellipses correspond to the $\Lambda$CDM- and blue ellipses to the modified gravity-model summarised in \autoref{tab:params}. The difference between both models are the fiducial values for the modified gravity and dark energy parameters. Both surveys use the same specification as described in \autoref{table:experimental_setup}. Note that the ellipses have been moved to the same point in parameter space for better comparison.}
\label{fig:compare_Fish_DEMG_LCDM}
\end{center}
\end{figure*}

In \autoref{fig:compare_Fish_DEMG_LCDM} we show the effect of the choice of the fiducial models. Red ellipses correspond to the $\Lambda$CDM, while the blue correspond to the modified gravity-model as described in \autoref{tab:params}. The main difference between both models is the background expansion due to the dark energy equation of state and the modified gravity parameterisation. It can be seen that the forecasted errors can change quite substantially. Most affected by this are the two dark energy parameters $w_0$ and $w_\mathrm{a}$, which can be understood quite well since the $w>-1$ supports clustering and thus increases the overall lensing signal relative to the noise. Moreover, the derivatives of the spectra tend to be larger if $w_\mathrm{a}>0$. One can also see a slight change of the ellipse in the $\alpha_\mathrm{B}-\alpha_\mathrm{M}$ plane which can be understood with the same reasoning. As pointed out in \citet{Schaefer2016}, the choice of the fiducial model is of great importance, especially for high dimensional parameter spaces. The difference found here can be of similar magnitude as effects due to marginalisation over nuisance parameters.

\citet{Alonso2017} investigated the constraints of stage IV surveys with a Fisher analysis only. To qualitatively compare both works we investigated a similar parameter space. A comparison with their best constraints \citet[Table 1 in][]{Alonso2017} shows that our constraints derived with the Fisher matrix formalism on modified gravity parameters are in general tighter. This effect is for example related to the marginalisation over $\alpha_\mathrm{T}$ and the modelling of intrinsic alignments. In contrast we find very similar results for the dark energy equation of state as well as for the neutrino masses. Generally we find good agreement with the results of \citet{Alonso2017}, especially with respect to parameter degeneracies. Since our main goal was the investigation of the likelihood with MCMC techniques, we refer the reader to \citet{Alonso2017} for a more detailed discussion on the influence of relativistic scales and alternative parameterisations.

\section{Conclusion}\label{sec:conclusion}
In this work we have investigated the potential of future cosmological surveys to constrain cosmological models based on scalar-tensor theories of the Horndeski class, including a dark energy and a neutrino component. The constraints draw information from a wide array of observations: Apart from temperature and polarisation anisotropies of the CMB and the reconstructed CMB lensing deflection field, we consider tomographic galaxy clustering and weak lensing to assemble the data covariance matrix and supply all possible non-zero cross-correlations. We assume that the modes of all large-scale structure probes follow Gaussian distributions, and that consequently the cosmological information is derived from two-point statistics. Error sources are instrumental noise in the case of the CMB, Poisson-noise for galaxy clustering, shape noise for weak lensing and the reconstruction noise of the CMB lensing deflection field. Apart from determining the parameter covariance in a Gaussian model for the likelihood through a Fisher matrix analysis we computed the likelihood through a Monte-Carlo Markov-chain, where we employed parallelised affine invariant sampling. The parameter space to be investigated by the MCMC sampler was only loosely restricted by priors and we assume the cosmological model to be spatially flat.

\begin{enumerate}
\item{Scalar-tensor theories of gravity as complex as the Horndeski class of models, can be well constrained by future surveys: Even the inclusion of a parameterised dark energy component affecting the FLRW-dynamics of the model and a neutrino component impacting on the amount of fluctuations on small spatial scales would allow a determination of the gravitational parameters with relative errors below 10\%. This is made possible by the combination of two-point correlation functions which are measured by stage-IV-surveys with significances in excess of $1400\sigma$ (CMB anisotropies), $750\sigma$ (galaxy clustering), $500\sigma$ (weak lensing) and $250\sigma$ (CMB-lensing), while strong cross-correlations, in particular between weak lensing and galaxy clustering, reduce the significance of the signal on one hand, but do in fact add new information and increase the sensitivity of the signal to the underlying cosmology on the other.}

\item{The phenomenology of scalar-tensor-theories include deviations of the propagation velocity of gravitational waves from the speed of light: as very tight bounds exist from the direct observation of gravitational waves, $\alpha_M$ is well constrained and would be an interesting value to compare with values resulting from structure formation. Alternatively, constraints on $\alpha_M$ from gravitational waves would be a strong prior for the cosmological model. In any case, the parameter constraining power of cosmological observations on $\alpha_B$ and $\alpha_M$, the dark energy component and the density parameters clearly originates from the combinations of probes at a wide range of redshifts and excludes variations of the gravitational constant over the age of the Universe that are larger than 10\%. Errors of that magnitude imply that cosmological surveys can complement Solar system tests of modified gravity models on large scales.}

\item{The constraints from gravitational wave experiments greatly reduce the volume in the theory space of the Horndeski class, as for example discussed in \citet{Amendola2017}. It will therefore be possible to constrain the remaining degrees of freedom at the $10\%$ level. The constraints derived by \citet{Alonso2017} are therefore slightly worse, since they did not use the constraints from gravitational waves.}

\item{Evaluation of the likelihoods over a 17-dimensional parameter space, which comprises gravity, the standard FLRW-parameters, the CDM-spectrum and astrophysical parameters such as galaxy bias or reionisation redshift, was carried out with a parallelised, affine-invariant MCMC-sampler. The MCMC-results were contrasted with those from a standard Fisher-matrix analysis, illustrating its limitations: While parameter degeneracies were usually approximated well by the Fisher-matrix, the magnitude of the statistical errors were usually underestimated by up to 30\% in the gravity parameters and the neutrino masses, by 50\% in the dark energy parameters and by a small margin in the standard parameters of a $\Lambda$CDM-model. It matters to a similar extent how the fiducial model is chosen, there are varying precisions to be expected if the true cosmology is a purely tensorial $\Lambda$CDM-model with a cosmological constant instead of dark energy or a scalar-tensor-theory, the latter leading to tighter constraints. In fact, this behaviour is known from clustering dark energy-cosmologies, where relaxing the value of the dark energy sound speed away from $c_\mathrm{s}=1$ enables clustering and leads to more precise measurements of $c_\mathrm{s}$ \citep{ayaita_investigating_2012}.}

\item{As it is difficult in cosmology to investigate certain aspects independently from others, we have included neutrinos, whose nonzero mass could be confirmed between $3\sigma$ and $4\sigma$, as well as a time-evolving dark energy component, where both the difference of the equation of state from $w=-1$ and its time evolution would be highly significant measurements. In a cosmology as complex as the type that we are investigating, many physical and statistical effects are relevant that determine the precision at which a parameter can be measured. Apart from a particular correlation function being directly sensitive to a parameter variation, there are indirect effects as well. For instance, the well-measured baryon density is a valuable input for the amplitude of baryon acoustic oscillations in galaxy clustering, which in turn improves constraints on dark energy through the background expansion history. For being able to quantify degeneracies between parameters, we have computed Pearson's correlation coefficient for each parameter pair in our 17-dimensional parameter space, from which we have identified the strongest degeneracies to be present in the group $\alpha_B$, $\alpha_M$ and $\alpha_T$, as they are all relevant in the evolution of gravitational potentials, in the group $w_0$ and $w_a$ because even this combination of probes can not constrain dark energy properties independently, and in the group $\Omega_\mathrm{m}$, $\sigma_8$, $h$ and $n_s$ as they determine jointly the shape and normalisation of the CDM-spectrum. One could, to some extent, expect further constraining power from the inclusion of non-Gaussian information, in the form of polyspectra \citep{hu_future_2012} or of cluster counts \citep{allen_cosmological_2011}, or by using new experimental techniques for large-scale structure surveys such as 21cm-cosmology \citep{hall_testing_2013, pourtsidou_testing_2015}.}

\item 
{\karl{It should be generally noted that the analysis presented here is far from final since we assumed several simplifying assumptions: (i) very simplistic model for galaxy clustering which ignores redshift errors, redshift space distortion and has a very simple model for the galaxy bias; (ii) simple noise curves for the CMB experiments at low multipoles and a vanishing cross-correlation between CMB lensing and the B-mode polarization; (iii) a simplistic description of non-linear structure growth; (iv) dereliction of intrinsic alignments in weak lensing maps.}}

\end{enumerate}

With these conclusions at hand, future prospects would involve the study of how systematic effects are be able to introduce biases, for instance how the intrinsic alignment effect tampers with gravitational lensing \citep{troxel_intrinsic_2015, joachimi_galaxy_2015, kiessling_galaxy_2015, kirk_galaxy_2015} or how the assumption of biasing models interferes with the galaxy clustering signal, how probes can be optimised to yield the best discrimination in model selection and how cosmological information flows from the observation to the model selection or to the final parameter constraint. Likewise, it would be interesting to see to what extent the larger, non-Gaussian error would change the model selection process and change Bayesian evidences for subclasses of Horndeski-cosmologies. In this respect, the true information content of non-linearly evolving and therefore non-Gaussian structures is difficult to quantify, but can be expected to add degeneracy-breaking power as well as further information on the phenomenology of gravity, in particular screening mechanisms. 

\section*{Acknowledgements}
RR and ASM acknowledge financial support from the graduate college {\em Astrophysics of cosmological probes of gravity} by Landesgraduiertenakademie Baden-W{\"u}rttemberg. We thank in particular Miguel Zumalacarregui for his help with \texttt{hi\char`_class}. We wish to thank an anonymous referee for his valuable comments to make the manuscript more clear.

\bibliographystyle{mnras}
\bibliography{MyBiB,old_MasterBib,spirou_mod_grav}

\begin{thebibliography}{}
\makeatletter
\relax
\def\mn@urlcharsother{\let\do\@makeother \do\$\do\&\do\#\do\^\do\_\do\%\do\~}
\def\mn@doi{\begingroup\mn@urlcharsother \@ifnextchar [ {\mn@doi@}
  {\mn@doi@[]}}
\def\mn@doi@[#1]#2{\def\@tempa{#1}\ifx\@tempa\@empty \href
  {http://dx.doi.org/#2} {doi:#2}\else \href {http://dx.doi.org/#2} {#1}\fi
  \endgroup}
\def\mn@eprint#1#2{\mn@eprint@#1:#2::\@nil}
\def\mn@eprint@arXiv#1{\href {http://arxiv.org/abs/#1} {{\tt arXiv:#1}}}
\def\mn@eprint@dblp#1{\href {http://dblp.uni-trier.de/rec/bibtex/#1.xml}
  {dblp:#1}}
\def\mn@eprint@#1:#2:#3:#4\@nil{\def\@tempa {#1}\def\@tempb {#2}\def\@tempc
  {#3}\ifx \@tempc \@empty \let \@tempc \@tempb \let \@tempb \@tempa \fi \ifx
  \@tempb \@empty \def\@tempb {arXiv}\fi \@ifundefined
  {mn@eprint@\@tempb}{\@tempb:\@tempc}{\expandafter \expandafter \csname
  mn@eprint@\@tempb\endcsname \expandafter{\@tempc}}}

\bibitem[\protect\citeauthoryear{Abazajian et~al.,}{Abazajian
  et~al.}{2016}]{CMBS4}
Abazajian K.~N.,  et~al., 2016, ArXiv e-prints 1610.02743

\bibitem[\protect\citeauthoryear{{Abbott} et~al.,}{{Abbott}
  et~al.}{2016}]{Ligo2016}
{Abbott} B.~P.,  et~al., 2016, \mn@doi [\prl] {10.1103/PhysRevLett.116.061102},
  \href {http://adsabs.harvard.edu/abs/2016PhRvL.116f1102A} {116, 061102}

\bibitem[\protect\citeauthoryear{{Abbott} et~al.,}{{Abbott}
  et~al.}{2017a}]{Abbott2017b}
{Abbott} B.~P.,  et~al., 2017a, \mn@doi [\prl]
  {10.1103/PhysRevLett.119.161101}, \href
  {http://adsabs.harvard.edu/abs/2017PhRvL.119p1101A} {119, 161101}

\bibitem[\protect\citeauthoryear{{Abbott} et~al.,}{{Abbott}
  et~al.}{2017b}]{Abbott2017a}
{Abbott} B.~P.,  et~al., 2017b, \mn@doi [\apjl] {10.3847/2041-8213/aa91c9},
  \href {http://adsabs.harvard.edu/abs/2017ApJ...848L..12A} {848, L12}

\bibitem[\protect\citeauthoryear{Achitouv, Baldi, Puchwein  \& Weller}{Achitouv
  et~al.}{2015}]{achitouv_imprint_2015}
Achitouv I.,  Baldi M.,  Puchwein E.,   Weller J.,  2015, ArXiv e-prints
  1511.01494

\bibitem[\protect\citeauthoryear{Acquaviva, Baccigalupi  \& Perrotta}{Acquaviva
  et~al.}{2004}]{acquaviva_weak_2004}
Acquaviva V.,  Baccigalupi C.,   Perrotta F.,  2004, \mn@doi [\prd]
  {10.1103/PhysRevD.70.023515}, 70

\bibitem[\protect\citeauthoryear{Adamek, Daverio, Durrer  \& Kunz}{Adamek
  et~al.}{2013}]{adamek_general_2013}
Adamek J.,  Daverio D.,  Durrer R.,   Kunz M.,  2013, \mn@doi [\prd]
  {10.1103/PhysRevD.88.103527}, 88

\bibitem[\protect\citeauthoryear{Albrecht et~al.,}{Albrecht
  et~al.}{2006}]{albrecht_report_2006}
Albrecht A.,  et~al., 2006, {ArXiv e-prints}

\bibitem[\protect\citeauthoryear{Allen, Evrard  \& Mantz}{Allen
  et~al.}{2011}]{allen_cosmological_2011}
Allen S.~W.,  Evrard A.~E.,   Mantz A.~B.,  2011, \mn@doi [Annual Review of
  Astronomy and Astrophysics] {10.1146/annurev-astro-081710-102514}, 49, 409

\bibitem[\protect\citeauthoryear{{Alonso}, {Bellini}, {Ferreira}  \&
  {Zumalac{\'a}rregui}}{{Alonso} et~al.}{2017}]{Alonso2017}
{Alonso} D.,  {Bellini} E.,  {Ferreira} P.~G.,   {Zumalac{\'a}rregui} M.,
  2017, \mn@doi [\prd] {10.1103/PhysRevD.95.063502}, \href
  {http://adsabs.harvard.edu/abs/2017PhRvD..95f3502A} {95, 063502}

\bibitem[\protect\citeauthoryear{{Amendola}, {Kunz}, {Saltas}  \&
  {Sawicki}}{{Amendola} et~al.}{2017}]{Amendola2017}
{Amendola} L.,  {Kunz} M.,  {Saltas} I.~D.,   {Sawicki} I.,  2017, preprint,
  \href {http://adsabs.harvard.edu/abs/2017arXiv171104825A} {} (\mn@eprint
  {arXiv} {1711.04825})

\bibitem[\protect\citeauthoryear{Ayaita, Sch{\"a}fer  \& Weber}{Ayaita
  et~al.}{2012}]{ayaita_investigating_2012}
Ayaita Y.,  Sch{\"a}fer B.~M.,   Weber M.,  2012, \mn@doi [{\mnras}]
  {10.1111/j.1365-2966.2012.20822.x}, 422, 3056

\bibitem[\protect\citeauthoryear{Babichev \& Deffayet}{Babichev \&
  Deffayet}{2013}]{babichev_introduction_2013}
Babichev E.,  Deffayet C.,  2013, \mn@doi [Classical and Quantum Gravity]
  {10.1088/0264-9381/30/18/184001}, 30, 184001

\bibitem[\protect\citeauthoryear{{Bailoni}, {Spurio Mancini}  \&
  {Amendola}}{{Bailoni} et~al.}{2017}]{Bailoni2017}
{Bailoni} A.,  {Spurio Mancini} A.,   {Amendola} L.,  2017, \mn@doi [\mnras]
  {10.1093/mnras/stx1209}, 470, 688

\bibitem[\protect\citeauthoryear{{Baldi} \& {Villaescusa-Navarro}}{{Baldi} \&
  {Villaescusa-Navarro}}{2016}]{Baldi2016}
{Baldi} M.,  {Villaescusa-Navarro} F.,  2016, preprint, \href
  {http://adsabs.harvard.edu/abs/2016arXiv160808057B} {} (\mn@eprint {arXiv}
  {1608.08057})

\bibitem[\protect\citeauthoryear{{Baldi}, {Villaescusa-Navarro}, {Viel},
  {Puchwein}, {Springel}  \& {Moscardini}}{{Baldi} et~al.}{2014}]{Baldi2014}
{Baldi} M.,  {Villaescusa-Navarro} F.,  {Viel} M.,  {Puchwein} E.,  {Springel}
  V.,   {Moscardini} L.,  2014, \mn@doi [\mnras] {10.1093/mnras/stu259}, \href
  {http://adsabs.harvard.edu/abs/2014MNRAS.440...75B} {440, 75}

\bibitem[\protect\citeauthoryear{{Barreira}, {Li}, {Hellwing}, {Baugh}  \&
  {Pascoli}}{{Barreira} et~al.}{2013}]{Barreira2013}
{Barreira} A.,  {Li} B.,  {Hellwing} W.~A.,  {Baugh} C.~M.,   {Pascoli} S.,
  2013, \mn@doi [\jcap] {10.1088/1475-7516/2013/10/027}, \href
  {http://adsabs.harvard.edu/abs/2013JCAP...10..027B} {10, 027}

\bibitem[\protect\citeauthoryear{{Bartelmann}}{{Bartelmann}}{2010}]{Bartelmann2010}
{Bartelmann} M.,  2010, \mn@doi [Reviews of Modern Physics]
  {10.1103/RevModPhys.82.331}, \href
  {http://adsabs.harvard.edu/abs/2010RvMP...82..331B} {82, 331}

\bibitem[\protect\citeauthoryear{{Bartelmann} \& {Schneider}}{{Bartelmann} \&
  {Schneider}}{2001}]{Bartelmann2001}
{Bartelmann} M.,  {Schneider} P.,  2001, \mn@doi [\physrep]
  {10.1016/S0370-1573(00)00082-X}, \href
  {http://adsabs.harvard.edu/abs/2001PhR...340..291B} {340, 291}

\bibitem[\protect\citeauthoryear{{Bartelmann}, {Fabis}, {Berg}, {Kozlikin},
  {Lilow}  \& {Viermann}}{{Bartelmann} et~al.}{2014}]{Bartelmann2014}
{Bartelmann} M.,  {Fabis} F.,  {Berg} D.,  {Kozlikin} E.,  {Lilow} R.,
  {Viermann} C.,  2014, preprint, \href
  {http://adsabs.harvard.edu/abs/2014arXiv1411.0806B} {} (\mn@eprint {arXiv}
  {1411.0806})

\bibitem[\protect\citeauthoryear{Bartelmann, Fabis, Kozlikin, Lilow, Dombrowski
   \& Mildenberger}{Bartelmann et~al.}{2017}]{Bartelmann2017}
Bartelmann M.,  Fabis F.,  Kozlikin E.,  Lilow R.,  Dombrowski J.,
  Mildenberger J.,  2017, \mn@doi [New Journal of Physics]
  {10.1088/1367-2630/aa7e6f}, 19, 083001

\bibitem[\protect\citeauthoryear{{Battye} \& {Pearson}}{{Battye} \&
  {Pearson}}{2012}]{Battye2012}
{Battye} R.~A.,  {Pearson} J.~A.,  2012, \mn@doi [\jcap]
  {10.1088/1475-7516/2012/07/019}, \href
  {http://adsabs.harvard.edu/abs/2012JCAP...07..019B} {7, 019}

\bibitem[\protect\citeauthoryear{{Battye} \& {Pearson}}{{Battye} \&
  {Pearson}}{2013}]{Battye2013}
{Battye} R.~A.,  {Pearson} J.~A.,  2013, \mn@doi [\prd]
  {10.1103/PhysRevD.88.061301}, \href
  {http://adsabs.harvard.edu/abs/2013PhRvD..88f1301B} {88, 061301}

\bibitem[\protect\citeauthoryear{{Baumgart} \& {Fry}}{{Baumgart} \&
  {Fry}}{1991}]{Baumgart1991}
{Baumgart} D.~J.,  {Fry} J.~N.,  1991, \mn@doi [\apj] {10.1086/170166}, \href
  {http://adsabs.harvard.edu/abs/1991ApJ...375...25B} {375, 25}

\bibitem[\protect\citeauthoryear{{Bellini} \& {Sawicki}}{{Bellini} \&
  {Sawicki}}{2014}]{Bellini2014}
{Bellini} E.,  {Sawicki} I.,  2014, \mn@doi [\jcap]
  {10.1088/1475-7516/2014/07/050}, \href
  {http://adsabs.harvard.edu/abs/2014JCAP...07..050B} {7, 050}

\bibitem[\protect\citeauthoryear{Bernardeau, Colombi, Gaztanaga  \&
  Scoccimarro}{Bernardeau et~al.}{2002}]{Bernardeau2002}
Bernardeau F.,  Colombi S.,  Gaztanaga E.,   Scoccimarro R.,  2002, Physics
  Reports, 367, 1

\bibitem[\protect\citeauthoryear{{Berti} et~al.,}{{Berti}
  et~al.}{2015}]{Berti2015}
{Berti} E.,  et~al., 2015, \mn@doi [Classical and Quantum Gravity]
  {10.1088/0264-9381/32/24/243001}, \href
  {http://adsabs.harvard.edu/abs/2015CQGra..32x3001B} {32, 243001}

\bibitem[\protect\citeauthoryear{Bertschinger \& Zukin}{Bertschinger \&
  Zukin}{2008}]{bertschinger_distinguishing_2008}
Bertschinger E.,  Zukin P.,  2008, \mn@doi [\prd] {10.1103/PhysRevD.78.024015},
  D78, 024015

\bibitem[\protect\citeauthoryear{Blazek, Mandelbaum, Seljak  \&
  Nakajima}{Blazek et~al.}{2012}]{blazek_separating_2012}
Blazek J.,  Mandelbaum R.,  Seljak U.,   Nakajima R.,  2012, JCAP, 2012, 041

\bibitem[\protect\citeauthoryear{Blazek, {MacCrann}, Troxel  \& Fang}{Blazek
  et~al.}{2017}]{blazek_beyond_2017}
Blazek J.,  {MacCrann} N.,  Troxel M.~A.,   Fang X.,  2017, ArXiv e-prints
  1708.09247

\bibitem[\protect\citeauthoryear{Boubekeur, Giusarma, Mena  \&
  Ram�rez}{Boubekeur et~al.}{2014}]{boubekeur_current_2014}
Boubekeur L.,  Giusarma E.,  Mena O.,   Ram�rez H.,  2014, \mn@doi [\prd]
  {10.1103/PhysRevD.90.103512}, D90, 103512

\bibitem[\protect\citeauthoryear{{Bouchet}, {Colombi}, {Hivon}  \&
  {Juszkiewicz}}{{Bouchet} et~al.}{1995}]{Bouchet1995}
{Bouchet} F.~R.,  {Colombi} S.,  {Hivon} E.,   {Juszkiewicz} R.,  1995, \aap,
  \href {http://adsabs.harvard.edu/abs/1995A%26A...296..575B} {296, 575}

\bibitem[\protect\citeauthoryear{{Buchert}}{{Buchert}}{1992}]{Buchert1992}
{Buchert} T.,  1992, \mnras, \href
  {http://adsabs.harvard.edu/abs/1992MNRAS.254..729B} {254, 729}

\bibitem[\protect\citeauthoryear{{Casas}, {Kunz}, {Martinelli}  \&
  {Pettorino}}{{Casas} et~al.}{2017}]{Casas2017}
{Casas} S.,  {Kunz} M.,  {Martinelli} M.,   {Pettorino} V.,  2017, preprint,
  \href {http://adsabs.harvard.edu/abs/2017arXiv170301271C} {} (\mn@eprint
  {arXiv} {1703.01271})

\bibitem[\protect\citeauthoryear{Catelan, Kamionkowski  \& Blandford}{Catelan
  et~al.}{2001}]{catelan_intrinsic_2001}
Catelan P.,  Kamionkowski M.,   Blandford R.~D.,  2001, MNRAS, 320, L7

\bibitem[\protect\citeauthoryear{{Chevallier} \& {Polarski}}{{Chevallier} \&
  {Polarski}}{2001}]{Chevallier2001}
{Chevallier} M.,  {Polarski} D.,  2001, \mn@doi [International Journal of
  Modern Physics D] {10.1142/S0218271801000822}, \href
  {http://adsabs.harvard.edu/abs/2001IJMPD..10..213C} {10, 213}

\bibitem[\protect\citeauthoryear{{Clifton}, {Ferreira}, {Padilla}  \&
  {Skordis}}{{Clifton} et~al.}{2012}]{Clifton2012}
{Clifton} T.,  {Ferreira} P.~G.,  {Padilla} A.,   {Skordis} C.,  2012, \mn@doi
  [\physrep] {10.1016/j.physrep.2012.01.001}, \href
  {http://adsabs.harvard.edu/abs/2012PhR...513....1C} {513, 1}

\bibitem[\protect\citeauthoryear{{Cole} et~al.,}{{Cole}
  et~al.}{2005}]{Cole2005}
{Cole} S.,  et~al., 2005, \mn@doi [\mnras] {10.1111/j.1365-2966.2005.09318.x},
  \href {http://adsabs.harvard.edu/abs/2005MNRAS.362..505C} {362, 505}

\bibitem[\protect\citeauthoryear{Collaboration et~al.,}{Collaboration
  et~al.}{2016}]{Planck15parameters}
Collaboration P.,  et~al., 2016, \mn@doi [\aap] {10.1051/0004-6361/201525830},
  594, A13

\bibitem[\protect\citeauthoryear{{Cooray} \& {Sheth}}{{Cooray} \&
  {Sheth}}{2002}]{Cooray2002}
{Cooray} A.,  {Sheth} R.,  2002, \mn@doi [\physrep]
  {10.1016/S0370-1573(02)00276-4}, \href
  {http://adsabs.harvard.edu/abs/2002PhR...372....1C} {372, 1}

\bibitem[\protect\citeauthoryear{{Copeland}, {Garousi}, {Sami}  \&
  {Tsujikawa}}{{Copeland} et~al.}{2005}]{Copeland2005}
{Copeland} E.~J.,  {Garousi} M.~R.,  {Sami} M.,   {Tsujikawa} S.,  2005,
  \mn@doi [\prd] {10.1103/PhysRevD.71.043003}, \href
  {http://adsabs.harvard.edu/abs/2005PhRvD..71d3003C} {71, 043003}

\bibitem[\protect\citeauthoryear{{Creminelli} \& {Vernizzi}}{{Creminelli} \&
  {Vernizzi}}{2017}]{Creminelli2017}
{Creminelli} P.,  {Vernizzi} F.,  2017, \mn@doi [Physical Review Letters]
  {10.1103/PhysRevLett.119.251302}, \href
  {http://adsabs.harvard.edu/abs/2017PhRvL.119y1302C} {119, 251302}

\bibitem[\protect\citeauthoryear{Croft \& Metzler}{Croft \&
  Metzler}{2000}]{croft_weak-lensing_2000}
Croft R.~A.,  Metzler C.~A.,  2000, ApJ, 545, 561

\bibitem[\protect\citeauthoryear{Deffayet, Gao, Steer  \& Zahariade}{Deffayet
  et~al.}{2011}]{Deffayet2011}
Deffayet C.,  Gao X.,  Steer D.~A.,   Zahariade G.,  2011, \mn@doi [Phys. Rev.
  D] {10.1103/PhysRevD.84.064039}, 84, 064039

\bibitem[\protect\citeauthoryear{Desjacques, Jeong  \& Schmidt}{Desjacques
  et~al.}{2016}]{desjacques_large-scale_2016}
Desjacques V.,  Jeong D.,   Schmidt F.,  2016, ArXiv e-prints 1611.09787

\bibitem[\protect\citeauthoryear{{Di Dio}, {Durrer}, {Marozzi}  \&
  {Montanari}}{{Di Dio} et~al.}{2016}]{DiDio2016}
{Di Dio} E.,  {Durrer} R.,  {Marozzi} G.,   {Montanari} F.,  2016, \mn@doi
  [\jcap] {10.1088/1475-7516/2016/01/016}, \href
  {http://adsabs.harvard.edu/abs/2016JCAP...01..016D} {1, 016}

\bibitem[\protect\citeauthoryear{Dossett, Ishak, Parkinson  \& Davis}{Dossett
  et~al.}{2015}]{dossett_constraints_2015}
Dossett J.~N.,  Ishak M.,  Parkinson D.,   Davis T.~M.,  2015, \mn@doi [\prd]
  {10.1103/PhysRevD.92.023003}, 92

\bibitem[\protect\citeauthoryear{{Ezquiaga} \& {Zumalac{\'a}rregui}}{{Ezquiaga}
  \& {Zumalac{\'a}rregui}}{2017}]{Ezquiaga2017}
{Ezquiaga} J.~M.,  {Zumalac{\'a}rregui} M.,  2017, \mn@doi [Physical Review
  Letters] {10.1103/PhysRevLett.119.251304}, \href
  {http://adsabs.harvard.edu/abs/2017PhRvL.119y1304E} {119, 251304}

\bibitem[\protect\citeauthoryear{{Feldman}, {Kaiser}  \& {Peacock}}{{Feldman}
  et~al.}{1994}]{Feldman1994}
{Feldman} H.~A.,  {Kaiser} N.,   {Peacock} J.~A.,  1994, \mn@doi [\apj]
  {10.1086/174036}, \href {http://adsabs.harvard.edu/abs/1994ApJ...426...23F}
  {426, 23}

\bibitem[\protect\citeauthoryear{{Ferraro}, {Sherwin}  \& {Spergel}}{{Ferraro}
  et~al.}{2015}]{Ferraro2015}
{Ferraro} S.,  {Sherwin} B.~D.,   {Spergel} D.~N.,  2015, \mn@doi [\prd]
  {10.1103/PhysRevD.91.083533}, \href
  {http://adsabs.harvard.edu/abs/2015PhRvD..91h3533F} {91, 083533}

\bibitem[\protect\citeauthoryear{{Font-Ribera}, {McDonald}, {Mostek}, {Reid},
  {Seo}  \& {Slosar}}{{Font-Ribera} et~al.}{2014}]{Ribera2014}
{Font-Ribera} A.,  {McDonald} P.,  {Mostek} N.,  {Reid} B.~A.,  {Seo} H.-J.,
  {Slosar} A.,  2014, \mn@doi [\jcap] {10.1088/1475-7516/2014/05/023}, \href
  {http://adsabs.harvard.edu/abs/2014JCAP...05..023F} {5, 023}

\bibitem[\protect\citeauthoryear{Foreman-Mackey, Hogg, Lang  \&
  Goodman}{Foreman-Mackey et~al.}{2013}]{foreman-mackey_emcee:_2013}
Foreman-Mackey D.,  Hogg D.~W.,  Lang D.,   Goodman J.,  2013, \mn@doi
  [Publications of the Astronomical Society of the Pacific Publications of the
  Astronomical Society of the Pacific Publications of the Astronomical Society
  of the Pacific] {10.1086/670067}, 125, 306

\bibitem[\protect\citeauthoryear{Forero-Romero, Contreras  \&
  Padilla}{Forero-Romero et~al.}{2014}]{forero-romero_cosmic_2014}
Forero-Romero J.~E.,  Contreras S.,   Padilla N.,  2014, \mn@doi [MNRAS]
  {10.1093/mnras/stu1150}, 443, 1090

\bibitem[\protect\citeauthoryear{Giannantonio, Martinelli, Silvestri  \&
  Melchiorri}{Giannantonio et~al.}{2010}]{giannantonio_new_2010}
Giannantonio T.,  Martinelli M.,  Silvestri A.,   Melchiorri A.,  2010, \jcap,
  2010, 030

\bibitem[\protect\citeauthoryear{{Goodman} \& {Weare}}{{Goodman} \&
  {Weare}}{2010}]{Goodman2010}
{Goodman} J.,  {Weare} J.,  2010, \mn@doi [Communications in Applied
  Mathematics and Computational Science, Vol.~5, No.~1, p.~65-80, 2010]
  {10.2140/camcos.2010.5.65}, \href
  {http://adsabs.harvard.edu/abs/2010CAMCS...5...65G} {5, 65}

\bibitem[\protect\citeauthoryear{Hall, Bonvin  \& Challinor}{Hall
  et~al.}{2013}]{hall_testing_2013}
Hall A.,  Bonvin C.,   Challinor A.,  2013, \mn@doi [\prd]
  {10.1103/PhysRevD.87.064026}, 87

\bibitem[\protect\citeauthoryear{{Hamimeche} \& {Lewis}}{{Hamimeche} \&
  {Lewis}}{2009}]{Hamimeche2009}
{Hamimeche} S.,  {Lewis} A.,  2009, \mn@doi [\prd]
  {10.1103/PhysRevD.79.083012}, \href
  {http://adsabs.harvard.edu/abs/2009PhRvD..79h3012H} {79, 083012}

\bibitem[\protect\citeauthoryear{{Heavens}}{{Heavens}}{2003}]{Heavens2003}
{Heavens} A.,  2003, \mn@doi [\mnras] {10.1046/j.1365-8711.2003.06780.x}, \href
  {http://adsabs.harvard.edu/abs/2003MNRAS.343.1327H} {343, 1327}

\bibitem[\protect\citeauthoryear{{Heavens} \& {Taylor}}{{Heavens} \&
  {Taylor}}{1995}]{Heavens1995}
{Heavens} A.~F.,  {Taylor} A.~N.,  1995, \mn@doi [\mnras]
  {10.1093/mnras/275.2.483}, \href
  {http://adsabs.harvard.edu/abs/1995MNRAS.275..483H} {275, 483}

\bibitem[\protect\citeauthoryear{Heavens, Refregier  \& Heymans}{Heavens
  et~al.}{2000}]{heavens_intrinsic_2000}
Heavens A.,  Refregier A.,   Heymans C.,  2000, \mnras, 319, 649

\bibitem[\protect\citeauthoryear{Heavens, Kitching  \& Verde}{Heavens
  et~al.}{2007}]{heavens_model_2007}
Heavens A.~F.,  Kitching T.~D.,   Verde L.,  2007, \mnras, 380, 1029

\bibitem[\protect\citeauthoryear{{Heitmann}, {White}, {Wagner}, {Habib}  \&
  {Higdon}}{{Heitmann} et~al.}{2010}]{Heitmann2010}
{Heitmann} K.,  {White} M.,  {Wagner} C.,  {Habib} S.,   {Higdon} D.,  2010,
  \mn@doi [\apj] {10.1088/0004-637X/715/1/104}, \href
  {http://adsabs.harvard.edu/abs/2010ApJ...715..104H} {715, 104}

\bibitem[\protect\citeauthoryear{{Hilbert}, {Hartlap}  \&
  {Schneider}}{{Hilbert} et~al.}{2011}]{Hilbert2011}
{Hilbert} S.,  {Hartlap} J.,   {Schneider} P.,  2011, \mn@doi [\aap]
  {10.1051/0004-6361/201117294}, \href
  {http://adsabs.harvard.edu/abs/2011A%26A...536A..85H} {536, A85}

\bibitem[\protect\citeauthoryear{{Hinshaw} et~al.,}{{Hinshaw}
  et~al.}{2013}]{Hinshaw2013}
{Hinshaw} G.,  et~al., 2013, \mn@doi [\apjs] {10.1088/0067-0049/208/2/19},
  \href {http://adsabs.harvard.edu/abs/2013ApJS..208...19H} {208, 19}

\bibitem[\protect\citeauthoryear{{Hirata} \& {Seljak}}{{Hirata} \&
  {Seljak}}{2003}]{Hirata2003}
{Hirata} C.~M.,  {Seljak} U.,  2003, \mn@doi [\prd]
  {10.1103/PhysRevD.68.083002}, \href
  {http://adsabs.harvard.edu/abs/2003PhRvD..68h3002H} {68, 083002}

\bibitem[\protect\citeauthoryear{{Hoekstra} \& {Jain}}{{Hoekstra} \&
  {Jain}}{2008}]{Hoekstra2008}
{Hoekstra} H.,  {Jain} B.,  2008, \mn@doi [Annual Review of Nuclear and
  Particle Science] {10.1146/annurev.nucl.58.110707.171151}, \href
  {http://adsabs.harvard.edu/abs/2008ARNPS..58...99H} {58, 99}

\bibitem[\protect\citeauthoryear{{Horndeski}}{{Horndeski}}{1974}]{Horndeski1974}
{Horndeski} G.~W.,  1974, \mn@doi [International Journal of Theoretical
  Physics] {10.1007/BF01807638}, \href
  {http://adsabs.harvard.edu/abs/1974IJTP...10..363H} {10, 363}

\bibitem[\protect\citeauthoryear{{Hu}}{{Hu}}{2002}]{Hu2002a}
{Hu} W.,  2002, \mn@doi [\prd] {10.1103/PhysRevD.66.083515}, \href
  {http://adsabs.harvard.edu/abs/2002PhRvD..66h3515H} {66, 083515}

\bibitem[\protect\citeauthoryear{{Hu} \& {Okamoto}}{{Hu} \&
  {Okamoto}}{2002}]{Hu2002}
{Hu} W.,  {Okamoto} T.,  2002, \mn@doi [\apj] {10.1086/341110}, \href
  {http://adsabs.harvard.edu/abs/2002ApJ...574..566H} {574, 566}

\bibitem[\protect\citeauthoryear{{Hu} \& {Sawicki}}{{Hu} \&
  {Sawicki}}{2007}]{Hu2007}
{Hu} W.,  {Sawicki} I.,  2007, \mn@doi [\prd] {10.1103/PhysRevD.76.064004},
  \href {http://adsabs.harvard.edu/abs/2007PhRvD..76f4004H} {76, 064004}

\bibitem[\protect\citeauthoryear{Hu, Liguori, Bartolo  \& Matarrese}{Hu
  et~al.}{2012}]{hu_future_2012}
Hu B.,  Liguori M.,  Bartolo N.,   Matarrese S.,  2012, ArXiv e-prints
  1211.5032

\bibitem[\protect\citeauthoryear{Jain \& Zhang}{Jain \&
  Zhang}{2008}]{jain_observational_2008}
Jain B.,  Zhang P.,  2008, \mn@doi [\prd] {10.1103/PhysRevD.78.063503}, 78,
  063503

\bibitem[\protect\citeauthoryear{Jing}{Jing}{2002}]{jing_intrinsic_2002}
Jing Y.~P.,  2002, \mn@doi [MNRAS] {10.1046/j.1365-8711.2002.05899.x}, 335, L89

\bibitem[\protect\citeauthoryear{Joachimi et~al.,}{Joachimi
  et~al.}{2015}]{joachimi_galaxy_2015}
Joachimi B.,  et~al., 2015, \mn@doi [Space Science Reviews]
  {10.1007/s11214-015-0177-4}, 193, 1

\bibitem[\protect\citeauthoryear{{Joyce}, {Jain}, {Khoury}  \&
  {Trodden}}{{Joyce} et~al.}{2015}]{Joyce2015}
{Joyce} A.,  {Jain} B.,  {Khoury} J.,   {Trodden} M.,  2015, \mn@doi [\physrep]
  {10.1016/j.physrep.2014.12.002}, \href
  {http://adsabs.harvard.edu/abs/2015PhR...568....1J} {568, 1}

\bibitem[\protect\citeauthoryear{Joyce, Lombriser  \& Schmidt}{Joyce
  et~al.}{2016}]{joyce_dark_2016}
Joyce A.,  Lombriser L.,   Schmidt F.,  2016, ArXiv e-prints 1601.06133

\bibitem[\protect\citeauthoryear{Kiessling et~al.,}{Kiessling
  et~al.}{2015}]{kiessling_galaxy_2015}
Kiessling A.,  et~al., 2015, \mn@doi [Space Science Reviews]
  {10.1007/s11214-015-0203-6}, 193, 67

\bibitem[\protect\citeauthoryear{Kirk et~al.,}{Kirk
  et~al.}{2015}]{kirk_galaxy_2015}
Kirk D.,  et~al., 2015, \mn@doi [Space Science Reviews]
  {10.1007/s11214-015-0213-4}, 193, 139

\bibitem[\protect\citeauthoryear{Kitching, Heavens  \& Das}{Kitching
  et~al.}{2014}]{kitching_3d_2014}
Kitching T.~D.,  Heavens A.~F.,   Das S.,  2014, ArXiv e-prints 1408.7052

\bibitem[\protect\citeauthoryear{{Kitching}, {Alsing}, {Heavens}, {Jimenez},
  {McEwen}  \& {Verde}}{{Kitching} et~al.}{2017}]{Kitching2017}
{Kitching} T.~D.,  {Alsing} J.,  {Heavens} A.~F.,  {Jimenez} R.,  {McEwen}
  J.~D.,   {Verde} L.,  2017, \mn@doi [\mnras] {10.1093/mnras/stx1039}, \href
  {http://adsabs.harvard.edu/abs/2017MNRAS.469.2737K} {469, 2737}

\bibitem[\protect\citeauthoryear{Knox}{Knox}{1995}]{knox_determination_1995}
Knox L.,  1995, \mn@doi [{\prd}] {10.1103/PhysRevD.52.4307}, 52, 4307

\bibitem[\protect\citeauthoryear{Kobayashi, Tashiro  \& Suzuki}{Kobayashi
  et~al.}{2010}]{kobayashi_evolution_2010}
Kobayashi T.,  Tashiro H.,   Suzuki D.,  2010, \mn@doi [\prd]
  {10.1103/PhysRevD.81.063513}, 81

\bibitem[\protect\citeauthoryear{Koyama}{Koyama}{2006}]{koyama_structure_2006}
Koyama K.,  2006, \mn@doi [\jcap] {10.1088/1475-7516/2006/03/017}, 3, 17

\bibitem[\protect\citeauthoryear{Koyama}{Koyama}{2016}]{koyama_cosmological_2016}
Koyama K.,  2016, \mn@doi [Reports on Progress in Physics]
  {10.1088/0034-4885/79/4/046902}, 79, 046902

\bibitem[\protect\citeauthoryear{Kunz \& Sapone}{Kunz \&
  Sapone}{2007}]{kunz_dark_2007}
Kunz M.,  Sapone D.,  2007, \mn@doi [\prl] {10.1103/PhysRevLett.98.121301}, 98,
  121301

\bibitem[\protect\citeauthoryear{{LSST Science Collaboration} et~al.,}{{LSST
  Science Collaboration} et~al.}{2009}]{LSST2009}
{LSST Science Collaboration} et~al., 2009, preprint, \href
  {http://adsabs.harvard.edu/abs/2009arXiv0912.0201L} {} (\mn@eprint {arXiv}
  {0912.0201})

\bibitem[\protect\citeauthoryear{Laszlo \& Bean}{Laszlo \&
  Bean}{2007}]{laszlo_non-linear_2007}
Laszlo I.,  Bean R.,  2007, ArXiv e-prints 0709.0307, 709

\bibitem[\protect\citeauthoryear{{Laureijs} et~al.,}{{Laureijs}
  et~al.}{2011}]{Laureijs2011}
{Laureijs} R.,  et~al., 2011, ArXiv e-prints, 1110.3193, \href
  {http://adsabs.harvard.edu/abs/2011arXiv1110.3193L} {}

\bibitem[\protect\citeauthoryear{Lesgourgues}{Lesgourgues}{2011}]{Lesgourgues2011}
Lesgourgues J.,  2011, preprint (\mn@eprint {} {1104.2932})

\bibitem[\protect\citeauthoryear{{Lewis}}{{Lewis}}{2008}]{Lewis2008}
{Lewis} A.,  2008, \mn@doi [\prd] {10.1103/PhysRevD.78.023002}, \href
  {http://adsabs.harvard.edu/abs/2008PhRvD..78b3002L} {78, 023002}

\bibitem[\protect\citeauthoryear{{Lewis} \& {Challinor}}{{Lewis} \&
  {Challinor}}{2006a}]{Lewis2006}
{Lewis} A.,  {Challinor} A.,  2006a, \mn@doi [\physrep]
  {10.1016/j.physrep.2006.03.002}, \href
  {http://adsabs.harvard.edu/abs/2006PhR...429....1L} {429, 1}

\bibitem[\protect\citeauthoryear{Lewis \& Challinor}{Lewis \&
  Challinor}{2006b}]{lewis_weak_2006}
Lewis A.,  Challinor A.,  2006b, \mn@doi [Physics Reports]
  {10.1016/j.physrep.2006.03.002}, 429, 1

\bibitem[\protect\citeauthoryear{{Li}, {Barreira}, {Baugh}, {Hellwing},
  {Koyama}, {Pascoli}  \& {Zhao}}{{Li} et~al.}{2013}]{Li2013}
{Li} B.,  {Barreira} A.,  {Baugh} C.~M.,  {Hellwing} W.~A.,  {Koyama} K.,
  {Pascoli} S.,   {Zhao} G.-B.,  2013, \mn@doi [\jcap]
  {10.1088/1475-7516/2013/11/012}, \href
  {http://adsabs.harvard.edu/abs/2013JCAP...11..012L} {11, 012}

\bibitem[\protect\citeauthoryear{{Limber}}{{Limber}}{1953}]{Limber1953}
{Limber} D.~N.,  1953, \mn@doi [\apj] {10.1086/145672}, \href
  {http://adsabs.harvard.edu/abs/1953ApJ...117..134L} {117, 134}

\bibitem[\protect\citeauthoryear{{Linder}, {Seng{\"o}r}  \& {Watson}}{{Linder}
  et~al.}{2016}]{Linder2016}
{Linder} E.~V.,  {Seng{\"o}r} G.,   {Watson} S.,  2016, \mn@doi [\jcap]
  {10.1088/1475-7516/2016/05/053}, \href
  {http://adsabs.harvard.edu/abs/2016JCAP...05..053L} {5, 053}

\bibitem[\protect\citeauthoryear{{Lombriser} \& {Taylor}}{{Lombriser} \&
  {Taylor}}{2015}]{Lombriser2015}
{Lombriser} L.,  {Taylor} A.,  2015, \mn@doi [\prl]
  {10.1103/PhysRevLett.114.031101}, \href
  {http://adsabs.harvard.edu/abs/2015PhRvL.114c1101L} {114, 031101}

\bibitem[\protect\citeauthoryear{Lue, Scoccimarro  \& Starkman}{Lue
  et~al.}{2004}]{lue_differentiating_2004}
Lue A.,  Scoccimarro R.,   Starkman G.,  2004, \prd, 69, 044005

\bibitem[\protect\citeauthoryear{{Maartens}, {Abdalla}, {Jarvis}, {Santos}  \&
  {SKA Cosmology SWG}}{{Maartens} et~al.}{2015}]{Maartens2015}
{Maartens} R.,  {Abdalla} F.~B.,  {Jarvis} M.,  {Santos} M.~G.,   {SKA
  Cosmology SWG} f.~t.,  2015, preprint, \href
  {http://adsabs.harvard.edu/abs/2015arXiv150104076M} {} (\mn@eprint {arXiv}
  {1501.04076})

\bibitem[\protect\citeauthoryear{Mackey, White  \& Kamionkowski}{Mackey
  et~al.}{2002}]{mackey_theoretical_2002}
Mackey J.,  White M.,   Kamionkowski M.,  2002, \mn@doi [MNRAS]
  {10.1046/j.1365-8711.2002.05337.x}, 332, 788

\bibitem[\protect\citeauthoryear{Mead, Peacock, Heymans, Joudaki  \&
  Heavens}{Mead et~al.}{2015}]{Mead2015}
Mead A.~J.,  Peacock J.~A.,  Heymans C.,  Joudaki S.,   Heavens A.~F.,  2015,
  \mn@doi [\mnras] {10.1093/mnras/stv2036}, 454, 1958

\bibitem[\protect\citeauthoryear{Merkel \& Schaefer}{Merkel \&
  Schaefer}{2017}]{merkel_parameter_2017}
Merkel P.~M.,  Schaefer B.~M.,  2017, \mn@doi [MNRAS] {10.1093/mnras/stx1044},
  469, 2760

\bibitem[\protect\citeauthoryear{{Nicolis}, {Rattazzi}  \&
  {Trincherini}}{{Nicolis} et~al.}{2009}]{Nicolis2009}
{Nicolis} A.,  {Rattazzi} R.,   {Trincherini} E.,  2009, \mn@doi [\prd]
  {10.1103/PhysRevD.79.064036}, \href
  {http://adsabs.harvard.edu/abs/2009PhRvD..79f4036N} {79, 064036}

\bibitem[\protect\citeauthoryear{{Okamoto} \& {Hu}}{{Okamoto} \&
  {Hu}}{2003}]{Okamoto2003}
{Okamoto} T.,  {Hu} W.,  2003, \mn@doi [\prd] {10.1103/PhysRevD.67.083002},
  \href {http://adsabs.harvard.edu/abs/2003PhRvD..67h3002O} {67, 083002}

\bibitem[\protect\citeauthoryear{{Perlmutter}, {Gabi}, {Goldhaber}, {Goobar},
  {Groom}, {Hook}, {Kim}  \& {et~al.}}{{Perlmutter}
  et~al.}{1997}]{Perlmutter1997}
{Perlmutter} S.,  {Gabi} S.,  {Goldhaber} G.,  {Goobar} A.,  {Groom} D.~E.,
  {Hook} I.~M.,  {Kim} A.~G.,   {et~al.} 1997, \apj, \href
  {http://adsabs.harvard.edu/abs/1997ApJ...483..565P} {483, 565}

\bibitem[\protect\citeauthoryear{{Perlmutter}, {Aldering}, {Goldhaber}  \& {et
  al.}}{{Perlmutter} et~al.}{1999}]{Perlmutter1999}
{Perlmutter} S.,  {Aldering} G.,  {Goldhaber} G.,   {et al.} 1999, \mn@doi
  [\apj] {10.1086/307221}, \href
  {http://adsabs.harvard.edu/abs/1999ApJ...517..565P} {517, 565}

\bibitem[\protect\citeauthoryear{Pettorino, Wintergerst, Amendola  \&
  Wetterich}{Pettorino et~al.}{2010}]{pettorino_neutrino_2010}
Pettorino V.,  Wintergerst N.,  Amendola L.,   Wetterich C.,  2010, \mn@doi
  [\prd] {10.1103/PhysRevD.82.123001}, 82

\bibitem[\protect\citeauthoryear{{Planck Collaboration XIII}}{{Planck
  Collaboration XIII}}{2015}]{Planck2015_XIII}
{Planck Collaboration XIII} 2015, ArXiv e-prints, 1502.01589, \href
  {http://adsabs.harvard.edu/abs/2015arXiv150201589P} {}

\bibitem[\protect\citeauthoryear{{Planck Collaboration} et~al.,}{{Planck
  Collaboration} et~al.}{2015}]{2015arXiv150201590P}
{Planck Collaboration} et~al., 2015, preprint, \href
  {http://adsabs.harvard.edu/abs/2015arXiv150201590P} {} (\mn@eprint {arXiv}
  {1502.01590})

\bibitem[\protect\citeauthoryear{{Planck Collaboration} et~al.,}{{Planck
  Collaboration} et~al.}{2016}]{Planck2016_XIII}
{Planck Collaboration} et~al., 2016, \mn@doi [\aap]
  {10.1051/0004-6361/201525830}, \href
  {http://adsabs.harvard.edu/abs/2016A\%26A...594A..13P} {594, A13}

\bibitem[\protect\citeauthoryear{Pourtsidou}{Pourtsidou}{2015}]{pourtsidou_testing_2015}
Pourtsidou A.,  2015, ArXiv e-prints 1511.05927

\bibitem[\protect\citeauthoryear{Pratten, Munshi, Valageas  \& Brax}{Pratten
  et~al.}{2016}]{pratten_3d_2016}
Pratten G.,  Munshi D.,  Valageas P.,   Brax P.,  2016, ArXiv e-prints
  1602.06711

\bibitem[\protect\citeauthoryear{{Raccanelli}, {Montanari}, {Bertacca},
  {Dor{\'e}}  \& {Durrer}}{{Raccanelli} et~al.}{2016}]{Raccanelli2016}
{Raccanelli} A.,  {Montanari} F.,  {Bertacca} D.,  {Dor{\'e}} O.,   {Durrer}
  R.,  2016, \mn@doi [\jcap] {10.1088/1475-7516/2016/05/009}, \href
  {http://adsabs.harvard.edu/abs/2016JCAP...05..009R} {5, 009}

\bibitem[\protect\citeauthoryear{{Redlich}, {Bolejko}, {Meyer}, {Lewis}  \&
  {Bartelmann}}{{Redlich} et~al.}{2014}]{Redlich2014}
{Redlich} M.,  {Bolejko} K.,  {Meyer} S.,  {Lewis} G.~F.,   {Bartelmann} M.,
  2014, \mn@doi [\aap] {10.1051/0004-6361/201424553}, \href
  {http://adsabs.harvard.edu/abs/2014A%26A...570A..63R} {570, A63}

\bibitem[\protect\citeauthoryear{Renk, Zumalac{\'a}rregui, Montanari  \&
  Barreira}{Renk et~al.}{2017}]{renk_galileon_2017}
Renk J.,  Zumalac{\'a}rregui M.,  Montanari F.,   Barreira A.,  2017, ArXiv
  e-prints 1707.02263

\bibitem[\protect\citeauthoryear{{Riess}, {Filippenko}, {Challis}  \& {et
  al.}}{{Riess} et~al.}{1998}]{Riess1998}
{Riess} A.~G.,  {Filippenko} A.~V.,  {Challis} P.,   {et al.} 1998, \mn@doi
  [\aj] {10.1086/300499}, \href
  {http://adsabs.harvard.edu/abs/1998AJ....116.1009R} {116, 1009}

\bibitem[\protect\citeauthoryear{{Riess}, {Strolger}, {Tonry}, {Casertano},
  {Ferguson}  \& {et al.}}{{Riess} et~al.}{2004}]{Riess2004}
{Riess} A.~G.,  {Strolger} L.-G.,  {Tonry} J.,  {Casertano} S.,  {Ferguson}
  H.~C.,   {et al.} 2004, \mn@doi [\apj] {10.1086/383612}, \href
  {http://adsabs.harvard.edu/abs/2004ApJ...607..665R} {607, 665}

\bibitem[\protect\citeauthoryear{{Riess}, {Strolger}, {Casertano}, {Ferguson},
  {Mobasher}, {Gold}, {Challis}  \& {et~al.}}{{Riess} et~al.}{2007}]{Riess2007}
{Riess} A.~G.,  {Strolger} L.-G.,  {Casertano} S.,  {Ferguson} H.~C.,
  {Mobasher} B.,  {Gold} B.,  {Challis} P.~J.,   {et~al.} 2007, \mn@doi [\apj]
  {10.1086/510378}, \href {http://adsabs.harvard.edu/abs/2007ApJ...659...98R}
  {659, 98}

\bibitem[\protect\citeauthoryear{{Sachs} \& {Wolfe}}{{Sachs} \&
  {Wolfe}}{1967}]{Sachs1967}
{Sachs} R.~K.,  {Wolfe} A.~M.,  1967, \mn@doi [\apj] {10.1086/148982}, \href
  {http://adsabs.harvard.edu/abs/1967ApJ...147...73S} {147, 73}

\bibitem[\protect\citeauthoryear{{Sakstein} \& {Jain}}{{Sakstein} \&
  {Jain}}{2017}]{Sakstein2017}
{Sakstein} J.,  {Jain} B.,  2017, \mn@doi [Physical Review Letters]
  {10.1103/PhysRevLett.119.251303}, \href
  {http://adsabs.harvard.edu/abs/2017PhRvL.119y1303S} {119, 251303}

\bibitem[\protect\citeauthoryear{Santos, Devi  \& Alcaniz}{Santos
  et~al.}{2016}]{santos_bayesian_2016}
Santos B.,  Devi N.~C.,   Alcaniz J.~S.,  2016, ArXiv e-prints 1603.06563

\bibitem[\protect\citeauthoryear{{Sch{\"a}fer} \& {Reischke}}{{Sch{\"a}fer} \&
  {Reischke}}{2016}]{Schaefer2016}
{Sch{\"a}fer} B.~M.,  {Reischke} R.,  2016, \mn@doi [\mnras]
  {10.1093/mnras/stw1221}, \href
  {http://adsabs.harvard.edu/abs/2016MNRAS.460.3398S} {460, 3398}

\bibitem[\protect\citeauthoryear{{Schneider} \& {Teyssier}}{{Schneider} \&
  {Teyssier}}{2015}]{Schneider2015}
{Schneider} A.,  {Teyssier} R.,  2015, \mn@doi [\jcap]
  {10.1088/1475-7516/2015/12/049}, \href
  {http://adsabs.harvard.edu/abs/2015JCAP...12..049S} {12, 049}

\bibitem[\protect\citeauthoryear{{Scoccimarro} \& {Frieman}}{{Scoccimarro} \&
  {Frieman}}{1999}]{Scoccimarro1999}
{Scoccimarro} R.,  {Frieman} J.~A.,  1999, \mn@doi [\apj] {10.1086/307448},
  \href {http://adsabs.harvard.edu/abs/1999ApJ...520...35S} {520, 35}

\bibitem[\protect\citeauthoryear{{Sellentin} \& {Heavens}}{{Sellentin} \&
  {Heavens}}{2017}]{Sellentin2017}
{Sellentin} E.,  {Heavens} A.~F.,  2017, preprint, \href
  {http://adsabs.harvard.edu/abs/2017arXiv170704488S} {} (\mn@eprint {arXiv}
  {1707.04488})

\bibitem[\protect\citeauthoryear{Sellentin, Heymans  \&
  Harnois-D{\'e}raps}{Sellentin et~al.}{2017}]{Sellentin2017b}
Sellentin E.,  Heymans C.,   Harnois-D{\'e}raps J.,  2017, preprint (\mn@eprint
  {} {1712.04923})

\bibitem[\protect\citeauthoryear{{Smith} et~al.,}{{Smith}
  et~al.}{2003}]{Smith2003}
{Smith} R.~E.,  et~al., 2003, \mn@doi [\mnras]
  {10.1046/j.1365-8711.2003.06503.x}, \href
  {http://adsabs.harvard.edu/abs/2003MNRAS.341.1311S} {341, 1311}

\bibitem[\protect\citeauthoryear{{Spurio Mancini}, Reischke, Pettorino,
  Sch{\"a}efer  \& Zumalac{\'a}rregui}{{Spurio Mancini}
  et~al.}{2018}]{SpurioMancini2018}
{Spurio Mancini} A.,  Reischke R.,  Pettorino V.,  Sch{\"a}efer B.~M.,
  Zumalac{\'a}rregui M.,  2018, preprint (\mn@eprint {} {1801.04251})

\bibitem[\protect\citeauthoryear{{Takahashi}, {Sato}, {Nishimichi}, {Taruya}
  \& {Oguri}}{{Takahashi} et~al.}{2012}]{Takahashi2012}
{Takahashi} R.,  {Sato} M.,  {Nishimichi} T.,  {Taruya} A.,   {Oguri} M.,
  2012, \mn@doi [\apj] {10.1088/0004-637X/761/2/152}, \href
  {http://adsabs.harvard.edu/abs/2012ApJ...761..152T} {761, 152}

\bibitem[\protect\citeauthoryear{{Tegmark}, {Taylor}  \& {Heavens}}{{Tegmark}
  et~al.}{1997}]{Tegmark1997}
{Tegmark} M.,  {Taylor} A.~N.,   {Heavens} A.~F.,  1997, \mn@doi [\apj]
  {10.1086/303939}, \href {http://adsabs.harvard.edu/abs/1997ApJ...480...22T}
  {480, 22}

\bibitem[\protect\citeauthoryear{{Thornton} et~al.,}{{Thornton}
  et~al.}{2016}]{Thornton2016}
{Thornton} R.~J.,  et~al., 2016, \mn@doi [\apjs] {10.3847/1538-4365/227/2/21},
  \href {http://adsabs.harvard.edu/abs/2016ApJS..227...21T} {227, 21}

\bibitem[\protect\citeauthoryear{Troxel \& Ishak}{Troxel \&
  Ishak}{2015}]{troxel_intrinsic_2015}
Troxel M.~A.,  Ishak M.,  2015, \mn@doi [Physics Reports]
  {10.1016/j.physrep.2014.11.001}, 558, 1

\bibitem[\protect\citeauthoryear{Tugendhat \& Schaefer}{Tugendhat \&
  Schaefer}{2017}]{Tugendhat2017}
Tugendhat T.~M.,  Schaefer B.~M.,  2017, arXiV

\bibitem[\protect\citeauthoryear{Vanderveld, Caldwell  \& Rhodes}{Vanderveld
  et~al.}{2011}]{vanderveld_second-order_2011}
Vanderveld R.~A.,  Caldwell R.~R.,   Rhodes J.,  2011, \mn@doi [\prl]
  {10.1103/PhysRevD.84.123510}, 84, 123510

\bibitem[\protect\citeauthoryear{Vanderveld, Mortonson, Hu  \&
  Eifler}{Vanderveld et~al.}{2012}]{vanderveld_testing_2012}
Vanderveld R.~A.,  Mortonson M.~J.,  Hu W.,   Eifler T.,  2012, \mn@doi [PRD]
  {10.1103/PhysRevD.85.103518}, 85, 103518

\bibitem[\protect\citeauthoryear{{Velten}, {Jim{\'e}nez}  \& {Piazza}}{{Velten}
  et~al.}{2017}]{Velten2017}
{Velten} H.,  {Jim{\'e}nez} J.~B.,   {Piazza} F.,  2017, in International
  Journal of Modern Physics Conference Series. p. 1760016 (\mn@eprint {arXiv}
  {1703.00307}), \mn@doi{10.1142/S2010194517600163}

\bibitem[\protect\citeauthoryear{White}{White}{2016}]{white_marked_2016}
White M.,  2016, ArXiv e-prints 1609.08632

\bibitem[\protect\citeauthoryear{{Winther} et~al.,}{{Winther}
  et~al.}{2015}]{Winther2015}
{Winther} H.~A.,  et~al., 2015, \mn@doi [\mnras] {10.1093/mnras/stv2253}, \href
  {http://adsabs.harvard.edu/abs/2015MNRAS.454.4208W} {454, 4208}

\bibitem[\protect\citeauthoryear{{Yoo}, {Fitzpatrick}  \& {Zaldarriaga}}{{Yoo}
  et~al.}{2009}]{Yoo2009}
{Yoo} J.,  {Fitzpatrick} A.~L.,   {Zaldarriaga} M.,  2009, \mn@doi [\prd]
  {10.1103/PhysRevD.80.083514}, \href
  {http://adsabs.harvard.edu/abs/2009PhRvD..80h3514Y} {80, 083514}

\bibitem[\protect\citeauthoryear{{Zel'Dovich}}{{Zel'Dovich}}{1970}]{ZelDovich1970}
{Zel'Dovich} Y.~B.,  1970, \aap, \href
  {http://adsabs.harvard.edu/abs/1970A%26A.....5...84Z} {5, 84}

\bibitem[\protect\citeauthoryear{Zhao, Pogosian, Silvestri  \& Zylberberg}{Zhao
  et~al.}{2009}]{zhao_searching_2009}
Zhao G.-B.,  Pogosian L.,  Silvestri A.,   Zylberberg J.,  2009, \mn@doi [\prd]
  {10.1103/PhysRevD.79.083513}, 79

\bibitem[\protect\citeauthoryear{{Zhao}, {Li}  \& {Koyama}}{{Zhao}
  et~al.}{2011}]{Zhao2011}
{Zhao} G.-B.,  {Li} B.,   {Koyama} K.,  2011, \mn@doi [\prd]
  {10.1103/PhysRevD.83.044007}, \href
  {http://adsabs.harvard.edu/abs/2011PhRvD..83d4007Z} {83, 044007}

\bibitem[\protect\citeauthoryear{Zumalacarregui, Bellini, Sawicki  \&
  Lesgourgues}{Zumalacarregui et~al.}{2016}]{Zumalacarregui2016}
Zumalacarregui M.,  Bellini E.,  Sawicki I.,   Lesgourgues J.,  2016, preprint
  (\mn@eprint {} {1605.06102})

\makeatother
\end{thebibliography}

\label{lastpage}

\end{document}